\documentclass[default,iicol]{sn-jnl}

\jyear{2021}%
\usepackage{times}
\usepackage{epsfig}
\usepackage{comment}
\usepackage{bbding}
\usepackage{amssymb}
\usepackage{pifont}

\usepackage{color}
\usepackage{adjustbox}

\theoremstyle{thmstyleone}%
\theoremstyle{thmstyletwo}%

\theoremstyle{thmstylethree}%

\begin{document}

\title[L]{Efficient Burst Raw Denoising with Variance Stabilization and Multi-frequency Denoising Network}

\author[1]{\fnm{Dasong} \sur{Li}}\email{dasongli@link.cuhk.edu.hk}
\equalcont{These authors contributed equally to this work.}
\author[1]{\fnm{Yi} \sur{Zhang}}\email{zhangyi@link.cuhk.edu.hk}
\equalcont{These authors contributed equally to this work.}

\author[2]{\fnm{Ka Lung} \sur{Law}}
\author[1,3]{\fnm{Xiaogang} \sur{Wang}}\email{xgwang@ee.cuhk.edu.hk}
\author*[2]{\fnm{Hongwei} \sur{Qin}}\email{qinhongwei@sensetime.com}

\author*[1,3]{\fnm{Hongsheng} \sur{Li}}\email{hsli@ee.cuhk.edu.hk}

\affil[1]{\orgdiv{Multimedia Laboratory}, \orgname{The Chinese University of Hong Kong}}
\affil[2]{\orgdiv{SenseTime Research}}
\affil[3]{\orgdiv{Centre for Perceptual and Interactive Intelligence Limited}}


\abstract{With the growing popularity of smartphones, capturing high-quality images is of vital importance to smartphones.
The cameras of smartphones have small apertures and small sensor cells, which lead to the noisy images in low light environment. 
Denoising based on a burst of multiple frames generally outperforms single frame denoising but with the larger compututional cost.
In this paper, we propose an efficient yet effective burst denoising system.
We adopt a three-stage design: noise prior integration, multi-frame alignment and multi-frame denoising. First, we integrate noise prior by pre-processing raw signals into a variance-stabilization space, which allows using a small-scale network to achieve competitive performance. Second, we observe that it is essential to adopt an explicit alignment for burst denoising, but it is not necessary to integrate an learning-based method to perform multi-frame alignment. Instead, we resort to a conventional and efficient alignment method and combine it with our multi-frame denoising network. At last, we propose a denoising strategy that processes multiple frames sequentially. Sequential denoising avoids filtering a large number of frames by decomposing multiple frames denoising into several efficient sub-network denoising. As for each sub-network, we propose an efficient multi-frequency denoising network to remove noise of different frequencies.
Our three-stage design is efficient and shows strong performance on burst denoising. Experiments on synthetic and real raw datasets demonstrate that our method outperforms state-of-the-art methods, with less computational cost. Furthermore, the low complexity and high-quality performance make deployment on smartphones possible.}




\maketitle

\section{Introduction} \label{sec1}

Mobile photography becomes increasingly popular because of the surging number of smartphones worldwide.
However, the raw images captured by low-cost sensors of mobile devices usually show heavy noise, especially in low-light environments. The noise augmented in the imaging processing pipeline would dramatically deteriorate the visual quality. To improve the imaging quality, capturing and denoising a burst of multiple raw frames becomes a common choice to remove unpleasant noise.
Therefore, burst raw denoising becomes an essential task in mobile photography.

Burst raw denoising aims at reproducing the underlying scene from a burst of noisy raw image frames. 
Classical efficient burst denoising methods \cite{vll, hasinoff2016burst} generally consist of some or all of the following three processing steps, including noise prior integration, multi-frame alignment, and multi-frame denoising. 
Raw images can be captured with different shooting parameters, such as exposure time and sensor gain, which lead to a wide range of noise levels with unstable variances. Integrating the noise prior helps the methods to be aware of the noise level of the input image so that they can handle different noise levels with different denoising strengths.
The multi-frame alignment stage tries to align multiple neighboring frames to the reference frame.
The representative methods include block matching \cite{hasinoff2016burst} and optical flow-based \cite{liu2014burstdenoising} methods. 
The final multi-frame denoising stage aggregates spatial and temporal information to reproduce the clean frame from the multiple frames with consideration of possible alignment failures. Classical fusion operations include collaborative filtering \cite{V-BM3d} and frequency domain fusion \cite{hasinoff2016burst}. 

Recently deep learning-based methods outperform traditional methods by improving the different processing stages. 
Kernel Prediction Network (KPN) and its variants \cite{MildenhallKPN18,MKPN,BPN} have been exploited to conduct multi-frame alignment and denoising jointly. 
 RViDeNet \cite{RViDeNet} and BDNet \citep{BDNet} improve the performance of multi-frame denoising by adopting Deformable Convolution \cite{DCN, DCN2} to achieve explicit alignment. However, less effects are paid to noise prior integration, which hampers the performance of burst raw denoising.
Moreover, most learning-based methods \cite{MildenhallKPN18, BDNet} generally require large memory consumption and high computational cost. 

In this paper, we improve the efficiency and effectiveness of burst denoising by improving all three steps:

\vspace{4pt}
\noindent\textbf{Noise Prior Integration.}
In traditional methods \cite{hasinoff2016burst, vll}, the variations of different noise levels of the input image make no difference to the computational complexity as the estimated noise variance is used as the coefficients of Wiener filter \cite{vll, hasinoff2016burst} to modulate the denoising strength. 
For deep learning-based methods \cite{MildenhallKPN18,zhang2017beyond}, they need to  tackle a wide range of noise levels with a single network. The denoising strengths need to be adaptively adjusted by explicitly or implicitly estimating noise variances \cite{MildenhallKPN18}. The network capacity needs to be large enough to handle the varying noise levels.

We first analyze the key factors causing unstable variances of raw images by modeling CMOS signals. 
Then we introduce a variance-stabilizing transformation to stabilize the varying variances caused by these factors. 
The follow-up denoising can be then performed in the variance-stabilization space. The proposed strategy allows using a single lightweight network to handle different noise levels.

\vspace{4pt}
\noindent \textbf{Multi-frame Alignment.}
The mainstreaming burst denoising methods utilize kernel prediction to achieve multi-frame alignment and multi-frame denoising jointly. Learning-based alignment \cite{wang2019edvr, BDNet} with Deformable Convolution \cite{DCN, DCN2} also show improved performance in burst denoising \cite{BDNet} or video denoising \cite{RViDeNet}.
In our experiments, it is shown that explicit alignment is essential to handle the global misalignment among multiple frames.
Furthermore, we find that learning-based alignment do not show competitive performance against conventional alignment but brings much more computational cost.
Therefore, we adopt a conventional alignment and combine it with learning-based multi-frame denoising.

Specifically, we adopt a coarse-to-fine alignment algorithm and process coarse and fine scales hierarchically. At the coarse scales, we use homography flow to achieve global alignment. At the fine scales, we adopt block-matching to refine the alignment results.
To make our denoising network adaptive to alignment results, our follow-up multi-frame denoising network is trained on multiple frames aligned by our alignment.
The proposed alignment strategy achieves competitive performance compared with learning-based alignment but has less computational cost.

\vspace{4pt}
\noindent \textbf{Multi-frame Denoising.} Processing a burst of frames simultaneously requires a large network, which increases the computational cost and memory significantly.
Therefore, it is difficult to deploy deep learning-based multi-frame denoising \cite{MildenhallKPN18, wang2019edvr} algorithms on mobile devices. 
We propose a sequential denoising strategy to process multiple frames in sequential manners.
The proposed denoising network first performs spatial denoising on the reference frame to produce an intermediate denoising result. Then, each neighboring frame is sequentially input into the network to gradually refine the intermediate result to integrate the multiple neighboring frames' temporal information. The denoising system adopts a novel and efficient multi-frequency network architecture to take advantages of the neural network's capability on eliminating high-frequency noise and to perform denoise at different frequencies.

We evaluate our proposed method on the KPN dataset \cite{MildenhallKPN18} and CRVD raw dataset \cite{RViDeNet}. Extensive experiments demonstrate the effectiveness of our proposed burst denoising approach.
In summary, our proposed method has the following contributions:
\begin{itemize}
    \item We propose an efficient burst denoising system by improving the three stages of the burst denoising framework, noise prior integration, multi-frame alignment, and multi-frame denoising.
    \item We analyze the variation of CMOS raw images and propose a comprehensive variance stabilization technique for learning-based denoising, which demonstrates its superior performance on raw burst denoising.
    \item We propose a multi-frame framework to integrate multiple frames' temporal information sequentially and a multi-frequency denoising network to handle noise of different frequencies effectively.
\end{itemize}

\section{Related Work}\label{sec2}

Burst raw denoising involves a complex processing pipeline that aggregates spatial and temporal information from multiple frames and should be capable of handling a wide range of noise levels. 
The related work about \textit{noise prior integration}, \textit{multi-frame alignment}, and \textit{multi-frame denoising} are discussed as follows.

\subsection{Noise Prior}
Given an observed intensity $x$ and its underlying clean intensity $x^{*}$, we have the following relation:
\begin{equation}
    x = x^{*} + n,
\end{equation} where $n$ is the noise. Additive white Gaussian noise is widely used in previous works \cite{DBLP:conf/cvpr/BuadesCM05, V-BM3d}. However, Gaussian distribution cannot represent the signal-dependent photon noise in CMOS sensors.
To approximate real noise of CMOS sensors, multiple types of noise are explored for noise modeling, such as Poisson-Gaussian distribution \cite{foi_PG,practical_signal_dependent}, heteroscedastic Gaussian distribution \cite{Heteroscedastic} and more complicated modelings \cite{eld, zhang2021rethinking}.

\vspace{3pt}
\noindent\textbf{Prior-based Traditional Methods.} 
Representative traditional denoising methods usually are based on different priors.
This category of methods include anisotropic diffusion, total variation denoising \cite{rudin1992nonlinear}, wavelet domain denoising \cite{DBLP:journals/tip/PortillaSWS03}, sparse coding \cite{DBLP:conf/iccv/MairalBPSZ09}, image self-similarity \cite{DBLP:conf/cvpr/BuadesCM05, V-BM3d} and etc.
Total variation denoising \cite{rudin1992nonlinear} uses the statistical characteristics of images to remove noise. Sparsity coding \cite{DBLP:conf/iccv/MairalBPSZ09} enforces sparsity in dictionary learning methods to learn over-complete dictionaries from clean images.
As an important prior, image self-similarity shows the excellent performance against other methods. NLM \cite{DBLP:conf/cvpr/BuadesCM05} and BM3d \cite{V-BM3d} explore the presence of similiar features or patterns in a non-local manner.
Although above models are limited due to the assumptions on the prior of spatially invariant noisy or clean images, they can be applied to real raw data when a generalized anscombe transformation \cite{GAT} is applied.

\vspace{3pt}
\noindent\textbf{Learning-based Noise Prior Integration.} 
With the development of convolution neural networks, end-to-end CNNs have achieve great success in handling real-world image denoising.
DnCNN proposes residual learning for Gaussian denoising. A series of methods \cite{zhang2018ffdnet, MildenhallKPN18, PMRID} proposed to integrate the noise prior of raw image into deep neural networks.
Based on Poisson-Gaussian distribution or Gaussian Mixture distribution, \cite{zhang2018ffdnet, MildenhallKPN18} approximate pixel-wise noise variance to deal with the different noise levels.  PMRID \cite{PMRID} proposes a linear transform to handle different noise variances in different sensor gains.
In this work, we propose a novel variance stabilization technique and perform learning-based denoising in the space transformed with stable variance.

\subsection{Multi-frame Alignment}
The most long-standing method for multi-frame alignment is based on optical flow \cite{optical_determining, optical_Lucas,optical_Baker,optical_menze, deepflow}. 
Hierarchical structures \cite{hasinoff2016burst, liu2014burstdenoising} are explored to improve the efficiency of conventional alignment.
Many learning-based alignment methods have been proposed in video-related tasks as the substitute for the conventional alignment.
Learning-based optical flow \cite{toflow} and deformable convolution \cite{wang2019edvr} have been exploited for video interpolation, video super-resolution and video denoising.
RViDeNet \cite{RViDeNet} proposed pre-denoising modules to denoise each frame before the deformable convolution alignment, which, however, increases the computational burden significantly.

\subsection{Multi-frame Denoising}
The multi-frame denoising aims at merging multiple frames with alignment errors to reproduce clean image. Collaborative filtering\cite{DBLP:journals/tip/MaggioniBFE12, DBLP:conf/ipas/MaggioniBFE11} and frequency domain fusion\cite{hasinoff2016burst, vll} are representative approaches.
The mainstreaming learning-based methods implement multi-frame denoising without explicit alignment.
KPN \cite{MildenhallKPN18} proposes kernel prediction network to jointly conduct multi-frame alignment and denoising. MPKN \citep{MKPN} extends single kernel prediction to multiple kernels prediction.
BPN \citep{BPN} proposes basis prediction networks for larger kernels.
FastDVDNet \cite{fastdvdnet} proposes two-step cascaded methods for efficient video denoising without alignment modules.

\section{Methodology}
Our raw burst denoising system adopts a three-stage design, which includes {\it noise prior integration}, {\it multi-frame alignment}, and {\it multi-frame denoising}. 
To produce a clean frame, $N$ raw frames in a burst are first transformed into the noise variance stabilized space via noise prior integration and then aligned by {multi-frame alignment}. The transformed and aligned multiple frames are input into the {multi-frame denoising} network to produce the clean frame corresponding to the key frame. 

\subsection{Noise Prior Integration}
Burst denoising in real-world scenarios needs to handle a wide range of noise levels.   
Most previous methods \cite{MildenhallKPN18, zhang2018ffdnet} integrate noise prior by using estimated per-pixel variance as the extra input. 
However, these networks still need to remove the noise with varying variance.
To improve both performance and efficiency, we would like to reduce the learning complexity of the network by eliminating the unstable variances.
We first discuss the key factors causing unstable noise variances of raw intensity values via formulating the noise of raw CMOS signals.
To eliminate the unstable variances of different noise levels, we use the variance-stabilizing techniques to eliminate the unstable noise variance caused by these factors. Further analysis is provided to reveal that stabilizing noise variances allows using lightweight networks to achieve effective denoising.

\subsubsection{Noise Modeling of CMOS Signals}
The raw data of CMOS signals contains two primary sources of noise: shot noise and read noise. Shot noise is produced as a Poisson process with a variance equal to signal level. The read noise, an approximately Gaussian process, is caused by the sensor readout effects. The raw data is usually modeled as Poisson-Gaussian distribution \cite{foi_PG,practical_signal_dependent}:
\begin{equation}
    x \sim \sigma_s \mathcal{P}\left(\frac{x^{*}}{\sigma_s}\right) +  \mathcal{N}(0, \sigma_r^2),
\end{equation}
where $x$ is noisy measurement of the true intensity $x^{*}$. Two noise parameters $\sigma_s$ and $\sigma_r$ change across different images as the sensor gain (ISO) changes.

The variance of the noisy measurement $x$ is formulated as
\begin{equation}
    \mathrm{Var}[x] = \sigma_s x^{*} + \sigma_r^2.
\end{equation}
For a fixed sensor, the sensor gain (ISO) is the only factor affecting $\sigma_s$ and $\sigma_r$. The connection between the sensor gain and noise parameters $\sigma_s, \sigma_r$ are shown in Appendix~\ref{secA0}.
Therefore the variance is affected by the sensor gain and underlying intensity $x^{*}$. 
When the sensor gain increases, the variance of each pixel at one image increases. When the sensor gain is fixed, different brightness shows different variances in the image.

\subsubsection{Variance Stabilization}
We propose to transform the pixel values into a new space to eliminate the varying variances.

Firstly, we eliminate the unstable variance caused by the sensor gain. 
The observed intensity $x$ and the underlying true intensity $\acute{x}$ are scaled by $\frac{1}{\sigma_s}$,
\begin{equation}
   \acute{x} = \frac{x}{\sigma_s} \ , \ \acute{x}^{*} = \frac{x^{*}}{\sigma_s}.
   \label{eq:acute_x1}
\end{equation}
With the above transformation, $\acute{x}$ becomes a Poisson variable corrupted by additive Gaussian noise of variance $\acute{\sigma}^2 = \frac{\sigma_r^2}{\sigma_s^{2}}$:
\begin{equation}
    \acute{x} = \mathcal{P}\left(\acute{x}^{*}\right) + \mathcal{N}\left(0, \acute{\sigma}^2 \right).
    \label{eq:acute_x}
\end{equation}
The variance of $\acute{x}$ thus becomes
\begin{equation}
    \mathrm{Var}[\acute{x}] = \acute{x}^{*} + \acute{\sigma}^2.
\end{equation}

Then we need to eliminate the unstable variance caused by
the signal-dependent property of the Poisson distribution, which indicates that different intensities within the same image have varying variances.

We generalize the Freeman-Tukey transformation \cite{tukey-freeman-transformation} to transform 
the Poisson-Gaussian distribution (Eq. \eqref{eq:acute_x}) to
\begin{equation}
    y = \sqrt{\acute{x} + \acute{\sigma}^2} + \sqrt{\acute{x} + 1 + \acute{\sigma}^2}, 
    \label{eq:acute_x2}
\end{equation}
which is a Gaussian distribution with a constant variance at different intensities, i.e.,

the transformed intensity values is contaminated by Gaussian noise with unit variance. For details of the interpretation in the transformed space, please refer to \cite{M2012Poisson, tukey-freeman-transformation}. Then the following denoising network perform denoising on this space.

\vspace{4pt}
\noindent\textbf{Inverse Transformation.}
We perform the algebraic inverse of Eq. \eqref{eq:acute_x1} and Eq. \eqref{eq:acute_x2} to map the denoising output back into the raw linear space,
\begin{equation}
    \text{Inv}(y) = 
    \left(\frac{y^{4} -2y^{2}+1}{4y^{2}} - \acute{\sigma}^2\right) \sigma_s.
    \label{eq:inverse}
\end{equation}

This pixel transformation facilitates the training and generality of the follow-up denoising network as the signal-dependent components of the noise are eliminated in the transformed intensity space.
The follow-up denoising network can be more effectively trained based on the transformed intensities with signal-independent Gaussian noise.

\subsubsection{Analysis}

We here discuss related methods on handling the problem of varying noise levels in denoising for comparison.
First, most learning-based denoising methods \cite{zhang2018ffdnet, MildenhallKPN18} estimate per-pixel variance map as 
\begin{equation}
    \mathrm{Var}[x] = \sigma_s \max(x,0) +  \sigma_r^2.
\end{equation}
However, the ideal noise estimation should be
\begin{equation}
    \mathrm{Var}[x] = \sigma_s \alpha x^{*} +  \sigma_r^2.
\end{equation}
Since the true intensity $x^{*}$ cannot be observed, using observed intensity $x$ to replace $x^{*}$ introduces the errors in noise estimation. Furthermore, these denoising networks still need to handle varying noise levels. In contrast, applying variance stabilization would avoid the errors of noise estimation and allows the denoising network to handle stable variance.

To stabilize the variance from sensor gain, PMRID \cite{PMRID} proposed a $k$-sigma transform
\begin{equation}
    f_{k}(x) = \frac{x}{\sigma_s} + \frac{\sigma_r^2}{\sigma_s^2}
    \label{eq:k_sigma}
\end{equation}
to transform the images into an ISO-invariant space.
This transformation only eliminates the unstable variance caused by the sensor gain but neglects unstable variance of the Poisson-Gaussian distribution. PMRID \cite{PMRID} can be considered as only the first step of our proposed stabilization.

For stabilization of the Poisson-Gaussian distribution, 
Generalized Anscombe Transformation (GAT) \cite{GAT} extended Anscombe transformation \cite{Anscombe} to stabilize the variance of Poisson-Gaussian distribution. 
In contrast, our proposed method extends Tukey-Freeman Transformation \cite{tukey-freeman-transformation} for stabilization of the Poisson-Gaussian distribution with simple first-order approximation \cite{GAT} provided in Appendix A. 
It is observed in our experiments that our variance-stabilization technique for learning-based denoising shows better performance than GAT \cite{GAT}.

\subsection{Multi-frame Alignment}
Given the multiple frames for denoising reference frame, it is natural to utilize frame alignment methods \cite{liu2014burstdenoising, hasinoff2016burst, vll} to align the frames before multi-frame denoising to optimally utilize neighboring frames' contextual information. 
In video reconstruction tasks, learning-based optical flow \cite{toflow} and Deformable Convolution \cite{wang2019edvr} have been explored to perform multi-frame alignment as a substitute for the conventional alignment methods.
However, it is not practical to deploy learning-based alignment on mobile processors because of its large amount of computational cost and running time.
We decide to buck the trend and resort to conventional alignment methods, to achieve multi-frame alignment in an efficient manner.
It is observed in our experiments (Section \ref{sec:ablation}) that conventional alignment and learning-based alignment actually show comparable denoising performance with learning-based denoising networks. 
\begin{figure*}
    \centering
    \includegraphics[width=1.0\textwidth]{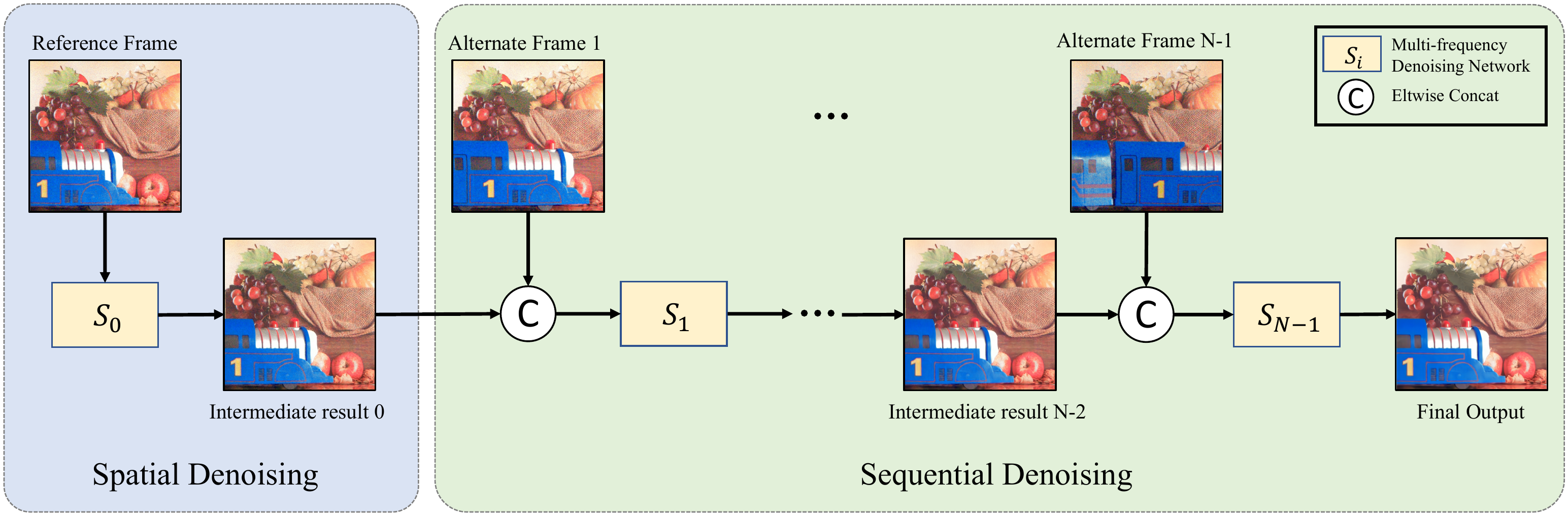}
    \caption{
    The overview of the proposed sequential denoising network. We propose a sequential denoising strategy to process the multiple frames. $N$ frames of a burst are considered as a reference frame and $N-1$ alternate frames. We adopt multi-frequency denoising network $S_i$ as the backbone for efficient denoising. At the first stage, we perform spatial denoising $S_0$ on the reference frame to obtain intermediate results 0. Then we refine the intermediate result via integrating the information from $N-1$ alternate frames. At each refinement stage, the multi-frequency denoising network $S_i$ takes intermediate result and an alternate frame as inputs and refines the intermediate result.} 
    \label{fig:overview}
\end{figure*}

\vspace{3pt}
\noindent\textbf{Coarse-to-fine Alignment.} Our alignment is based on the combination of block-matching \cite{hasinoff2016burst} and homography flow \cite{liu2014burstdenoising}. We build a four-scale alignment pyramid to estimate the motions between pairs of frames in a coarse-to-fine manner. We use homography flow at the top two levels and use block-matching for the bottom pyramid levels.
The homgraphy flow \cite{liu2014burstdenoising} is used to model accurate camera motions.
The two-layer homography flow acts as cascaded global alignment, while the block matching acts as refining local alignment. The cascade design makes our alignment both robust and fast under extreme low light environments.

The detailed steps of pyramid alignment is conducted as follows:
\begin{itemize}
  \item At the top pyramid level, we estimate global homography between the reference frame and other frames. All pixels share the same flow vector. The homography is calculated based on corner keypoints detected by FAST \cite{Fusing_points, high_speed_corner_detection} and feature matching with BRIEF descriptors \cite{BRIEF}. 
  
  \item At the second level, the whole plane is split into 4 blocks. Then we estimate the homography flow for each block separately. The BRIEF descriptors are extracted again from each block. When a block has insufficient matched features, the estimated homography would be substituted by the homography estimated from the previous level.
  
  \item At the two bottom levels, we use tile-based search to align frames. 
  The pre-aligned images are split into $16 \times 16$  tiles. We perform $L1$-distance search for each tile within $\pm$2 pixels neighborhood. In extreme low light case, tile size would be set to $32 \times 32$ to reduce the impact of noises. We accelerate the $L1$ search on ARM Neon Intrinsics \cite{Arm}. 
\end{itemize}
After the above alignment, there still remains some misalignments caused by the following reasons. 1) The large displacements would cause inevitable misalignments. 2) Pixels in one tile are required to share the same motion vector, which causes the unsmoothed translations between neighboring tiles. 3) We remove the operation of sub-pixel translation in \cite{hasinoff2016burst} to avoid interpolation. 
To handle these misalignments, we train our follow-up multi-frame denoising network on images aligned by our proposed alignment method.

\subsection{Multi-frame Denoising}
\begin{figure*}
    \centering
    \includegraphics[width=1.0\textwidth]{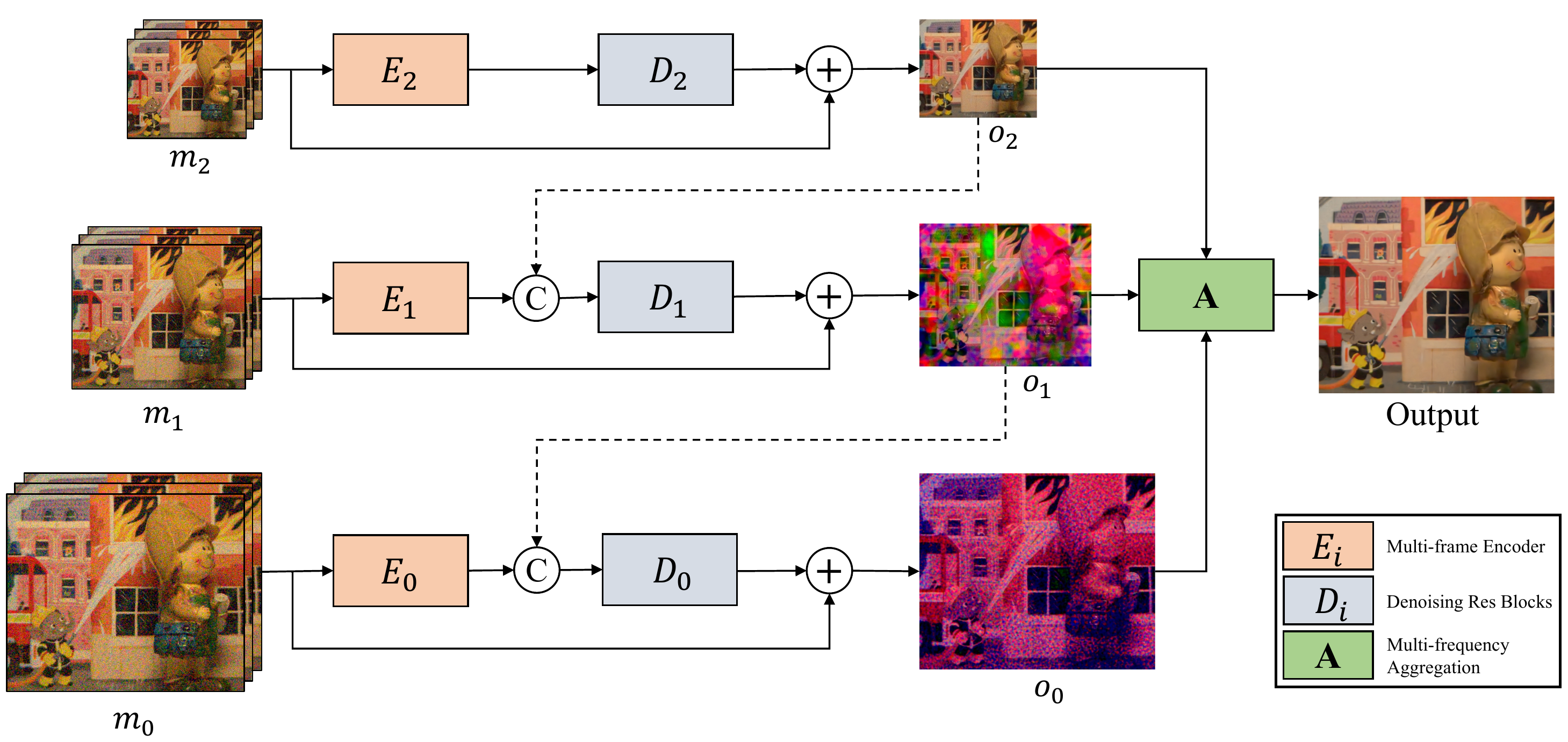}
    \caption{The overview of multi-frequency denoising network. The noises of different frequencies are separately handled by the multi-frequency denoising network.
    The network adopts three scales to handle different frequencies. The input of three scales are $m_2$, $m_1$, $m_0$, where $m_2$ and $m_1$ are downsampling versions of $m_0$.
    The output $o_2$, $o_1$, $o_0$ of three scales contain noises of different structures. At last, Multi-frequency Aggregation takes and aggregates $o_0$, $o_1$, $o_2$ to remove noise of all frequencies and preserve as many details as possible. The input $m_0,m_1,m_2$ and output $o_1,o_2,o_3$ are in raw Bayer pattern. For visualization, we utilize the post-process pipeline in RViDeNet \cite{RViDeNet} to process them into the sRGB space.
    }
    \label{fig:frequency}
\end{figure*}

After the multiple frames in a burst are aligned with the above stage, the multi-frame denoising stage needs to aggregate temporal information from the multiple frames to produce one clean frame. 
However, processing a large number of frames \cite{MildenhallKPN18,fastdvdnet, RViDeNet} simultaneously needs to adopt heavy networks as the relations between the too many frames might be challenging to model. 
To mitigate the need of heavy networks, we process multiple frames sequentially by a series of efficient networks. All networks shares the multi-frequency denoising architecture with different parameters.
The overview of our proposed multi-frame denoising network is shown in Fig.~\ref{fig:frequency}. In Section~\ref{ssec:sequantialdenoising}, we introduce our sequential denoising strategy. In Section \ref{ssec:multiscalesubnetwork}, we present the proposed multi-frequency denoising network.

\subsubsection{Sequential denoising}
\label{ssec:sequantialdenoising}
For the input $N$ frames, one of them is selected as the reference frame and the others $N$-1 are the alternate frames.
The denoising network $S$ consists of $N$ sub-networks $S_0, S_1, \dots, S_{N-1}$ that process these frames sequentially.

\vspace{4pt}
\noindent\textbf{Stage-1: Single-frame denoising of the reference frame.} 
The multi-frame denoising network is trained to produce one clean image with the same content of the reference frame. 
Intuitively, the reference frame generally makes more contributions than alternate frames in the input clip for multi-frame denoising. 
We process the reference frame separately as the first stage of denoising. Our first sub-network $S_0$ performs single-frame denoising on the reference frame to produce an intermediate result $I_0$, which contains the same content with the noisy reference frame but has sharper edges and less noise.
Performing the single-frame pre-denoising on the reference frame as the first step offers the following benefits:
1) Processing the reference frame separately emphasizes the priority of reference frame over other alternate frames and generates an intermediate result.
It is observed in our experiments that refining the intermediate result $I_0$ shows better performance than directly refining noisy reference frame with alternate frames directly. 
2) The first stage only performs single-frame denoising on the reference frame and does not need consider the temporal relations between frames. Hence only a lightweight network can be adopted for efficiency. This strategy improves the following multi-frame denoising significantly but with less computational cost.

\vspace{4pt}
\noindent\textbf{Stage-2: Multi-frame denoising of the reference frame.}
Even with the alignment, the alternate frames might still show various degrees of misalignment to the reference frame caused by occlusions, blurring and sub-pixel translations. Therefore, different alternate frames should have different contributions for denoising the reference frame.
To avoid too much extra computational cost, our system processes $N-1$ alternate frames sequentially.
The intermediate result $I_0$, produced by the single-frame denoising of the reference frame, is refined sequentially by the $N-1$ separate sub-networks, $S_1, S_2, \dots, S_{N-1}$.
Specifically, $S_i$ takes the intermediate result $I_{i-1}$ from the previous sub-network and the $i$-th group of alternate frames as inputs, and generate the refined intermediate result $I_i$.
The intermediate result $I_{N-1}$ of the last sub-network $S_{N-1}$ generates the final denoising output.

This proposed sequential denoising strategy is quite efficient as it avoids handling all frames jointly and only processes one alternate frame at a time. A series of lightweight sub-networks can be adopted. But the proposed method can still incorporate all information of the neighboring frames to denoise the reference frame gradually.

\subsubsection{Multi-frequency Denoising Network}
\label{ssec:multiscalesubnetwork}
Each of the denoising sub-network $S_1, \dots, S_{N-1}$ takes the intermediate result and one alternate frame
as input and produces the clean frame corresponding to the reference frame. 
As observed by \cite{disneyasy}, denoising neural networks are typically good at removing high-frequency noise but have more difficulties on handling low-frequency noise.
Therefore, we decompose denoising of whole images into denoising on multiple frequencies. 
Our sub-network consists of a \textit{multi-frequency denoising} and a \textit{multi-frequency aggregation} module. 

\noindent\textbf{Multi-frequency Denoising.} 
We first perform bilinear downsampling on the input frames by a factor of 2 for two times to obtain three-scale image pyramids $\{m^0, m^1, m^2\}$. 
Noise of the same pattern would correspond to different frequencies when downsampled to different scales. For instance, the low-frequency noise would be of high-frequency if the image is downsampled. 
Then we perform denoising at the three scales to remove noise of different frequencies.

At scale $i$, the denoising sub-network $F_i$, containing a multi-frame encoder $E^i$ and a denoising decoder $D^i$, predicts the residual noise for the image $m^i$. The multi-frame encoder $E^i$ is implemented by three-layer Convolution neural network. The denoising decoder $D^i$ utilize four residual blocks \cite{Resnet}.
The intermediate denoised images $o^i$ is obtained as
 \begin{align}
  o^i &= D^i(E^i(m^i)) + m^i.
\end{align}

Inspired by previous image restoration networks \citep{SGN_2019_ICCV, GoPro}, we further propagate the intermediate result $o_i$ of lower frequency to the decoder $D^{i-1}$ for handling higher-frequency noise. 
The intermediate denoised images $o_i$ are calculated as
\begin{align}
  o^2 &= D^2(E^2(m^2)) + m^2, \\
  o^1 &= D^1(E^1(m^1), o^2) + m^1, \\
  o^0 &= D^0(E^0(m^0), o^1) + m^0. 
\end{align}
In our network, $D^i$ takes the encoder feature $E^i(m^i)$ and the intermediate denoising result $o^{i+1}$ from the lower frequency as inputs and generates the intermediate result at scale $i$. Intuitively, $D^2$ works on the smallest-scale image and the low-frequency noise in the original image is mostly removed. $D^1$ takes the intermediate result from $D^2$ and further conducts denoising in a higher range of frequency, and $D^3$ repeats this procedure to work on noise of even higher frequency. In this way, noise of different frequencies are efficiently handled by different sub-networks.

\begin{table*}[!t]
\footnotesize
\centering
\begin{tabular}{lccccc}
\toprule
Method      & Gain $\propto$ 1 & Gain $\propto$ 2 & Gain $\propto$ 4 & Gain $\propto$ 8 & Average \\
\midrule
Noisy ref.  &   28.70 & 24.19 &  19.80 &  15.76 & 22.11 \\ 
V-BM4D \cite{DBLP:conf/ipas/MaggioniBFE11} & 34.60 &  31.89 & 29.20 &  26.52 &  30.55 \\ 
KPN \cite{MildenhallKPN18}  &  36.47 & 33.93  & 31.19  & 27.97 & 32.39 \\
MKPN \cite{MKPN} &  36.88 & 34.22  & 31.45  & 28.52 & 32.77 \\ 
BPN \cite{BPN}      & 38.18  & 35.42 & 32.54 & 29.45 & 33.90 \\
\hline
Ours & \textbf{39.39} & \textbf{36.52} & \textbf{33.47} & 
\textbf{30.20} & \textbf{34.90} \\
\bottomrule
\end{tabular}
\caption{The PSNR results of different burst denoising methods on KPN dataset.}
\label{result_kpn}
\footnotesize
\vspace{4pt}
\begin{tabular}{ccccccccc}
\toprule
 &  & Noisy Input & V-BM4D \cite{V-BM3d} & EMVD \cite{EMVD} &  FastDVDNet \cite{fastdvdnet} &  RViDeNet \cite{RViDeNet} & ours  \\ 
 \midrule
\multirow{2}{*}{$N=3$} & raw & 32.01 / 0.732 & -  & 44.05 / 0.989 & - & 44.08 / 0.988 & \textbf{44.43} / \textbf{0.989} \\
~ & sRGB & 31.79 / 0.752 & 35.20 / 0.9577 & 39.53 / 0.978 & - & 40.03 / 0.980 & \textbf{40.49} / \textbf{0.982} \\ \hline
\multirow{2}{*}{$N=5$} & raw & 32.01 / 0.732 & - & 44.05 / 0.989 & 44.30 / 0.989 & 44.30 / 0.988 & \textbf{44.70} / \textbf{0.990} \\
~ & sRGB & 31.79 / 0.752 & -  & 39.53 / 0.978 & 40.27 / 0.981 & 40.27 / 0.981 & \textbf{40.88} / \textbf{0.983} \\ 
\bottomrule
\end{tabular}
\caption{The PSNR and SSIM results of different burst denoising methods on CRVD dataset. }
\label{table_crvd}
\end{table*} 
\vspace{4pt}
\noindent\textbf{Multi-frequency Aggregation.}
After performing multi-frequency denoising, we design a lightweight multi-frequency aggregation module to combine the denoising results $o^0, o^1, o^2$ of the three scales to generate the final output.
To capture low frequency noise of $o^0$ and $o^1$, the noise residuals are calculated
\begin{equation}
    n^{1} = \downarrow(o^0) - o^{1}, \quad
	n^{2} = \downarrow(o^1) - o^{2}.
\end{equation}
Intuitively, $o^0$ has its high-frequency noise at scale $0$ removed. To further capture its low-frequency noise, it is downsampled to scale $0$ as $\downarrow(o^0)$, so that its remaining low-frequency noise can be converted to high-frequency noise at scale $1$. Since $o^1$ has the high-frequency noise at scale $1$ removed, the residual $n^{1} = \downarrow(o^0) - o^{1}$ highlights the lower-frequency noise in $o^0$. Similarly, $n^2$ captures even lower-frequency noise of $o^0$ than $n^1$. In Fig \ref{fig:frequency}, we illustrate the gradually removed noise from high frequency to low frequency.
The captured lower-frequency noise $n^1$ and $n^2$ can then be removed at scale $0$ to obtain the refined output image $I$,
\begin{align}
	I = o^0 - \uparrow(n^1) - \uparrow(\uparrow(n^2)).
\end{align}
Since $n^1$ and $n^2$ have smaller size than $o^0$, they are upsampled by a factor of 2 once and twice respectively to match the size of $o^0$.
The proposed denoising network conducts denoising at different frequencies and achieves optimal performance in a multi-scale manner.

\noindent\textbf{Loss function.} We optimize our denoising network $S$ in the space of variance stabilization. Based on Eq.~\eqref{eq:acute_x2}, the network input $y$ is in the space of variance stabilization and the denoising output is denoted as $\hat{y} = S(y)$.
The ground truth $y^{*}$ in this space is obtained as
\begin{equation}
    y^{*} = \sqrt{\acute{x}^* + \acute{\sigma}^2} + \sqrt{\acute{x}^* + 1 + \acute{\sigma}^2},
\end{equation}
where $\acute{x}^{*}$ is obtained via Eq.~\eqref{eq:acute_x1}.
We use the average $\mathcal{L}_1$ distance and gradient loss \cite{MildenhallKPN18} as the main loss term.
Our loss function can be formulated as
\begin{align}
	\mathcal{L}_{r} = \mathcal{L}_1(y^{*}, \hat{y}) + w_1 \mathcal{L}_1(\nabla y^{*}, \nabla \hat{y}),
	\label{loss_1}
\end{align}
where $\nabla$ is the finite difference operator that convolves its input with $\left[-1,1\right]$ and $\left[-1,1\right]^{T}$, and $w_1$ is set to 0.5 in our experiments.

\noindent When we have an ISP to process the raw output to sRGB domain for assessment, we add the $\mathcal{L}_1$ distance in the sRGB domain as \cite{RViDeNet} does.
The loss function then becomes
\begin{align} \label{eq:loss_2}
    \mathcal{L}_{r} = & \mathcal{L}_1(y^{*}, \hat{y}) 
    + w_1  \mathcal{L}_1(\nabla y^{*}, \nabla \hat{y}) + \\ \nonumber
    & w_2 \mathcal{L}_1(\text{ISP}(\text{Inv}(y^{*})), \text{ISP}(\text{Inv}(\hat{y}))),
\end{align}
where $\text{Inv}(y^{*})$ and $\text{Inv}(\hat{y})$ follows Eq. \eqref{eq:inverse}, and $w_2$ is set to 0.5.

\section{Experiments}

We first evaluate the overall performance of our proposed method on the KPN dataset \cite{MildenhallKPN18} and raw video benchmarks CRVD dataset \cite{RViDeNet}.  
We compare our method against state-of-the-art burst and video denoising methods, including VBM4D \cite{DBLP:conf/ipas/MaggioniBFE11} FastDVDNet \cite{fastdvdnet}, KPN \cite{MildenhallKPN18} and RViDeNet \cite{RViDeNet}. 

To evaluate specific designs of the three modules, we conduct ablation study to investigate the influence of each module. Finally, we present our methods's actual deployment and inference speed on smartphones on a Snapdragon 888 processor \cite{processor_888}.

\setlength{\tabcolsep}{2pt}
\begin{figure*}[!t]
    \small
    \begin{center}
    \begin{tabular}{cccccc}
        \includegraphics[width=.221\linewidth]{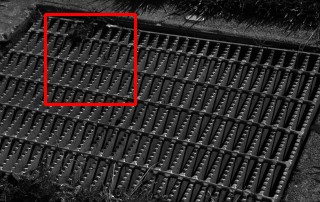} &
        \includegraphics[width=.14\linewidth]{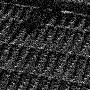} &
        \includegraphics[width=.14\linewidth]{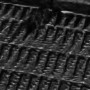} &
        \includegraphics[width=.14\linewidth]{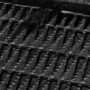} &
        \includegraphics[width=.14\linewidth]{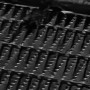} &
        \includegraphics[width=.14\linewidth]{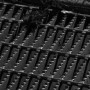} \\
        \includegraphics[width=.221\linewidth]{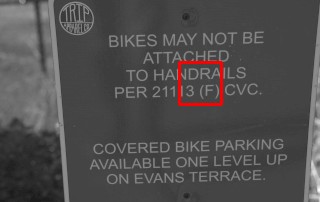} &
        \includegraphics[width=.14\linewidth]{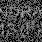} &
        \includegraphics[width=.14\linewidth]{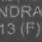} &
        \includegraphics[width=.14\linewidth]{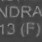} &
        \includegraphics[width=.14\linewidth]{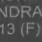} &
        \includegraphics[width=.14\linewidth]{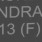} \\
        \includegraphics[width=.221\linewidth]{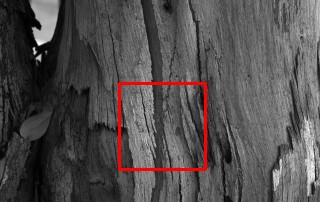} &
        \includegraphics[width=.14\linewidth]{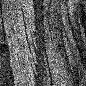} &
        \includegraphics[width=.14\linewidth]{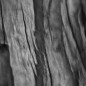} &
        \includegraphics[width=.14\linewidth]{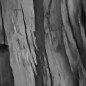} &
        \includegraphics[width=.14\linewidth]{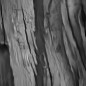} &
        \includegraphics[width=.14\linewidth]{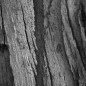} \\
        \includegraphics[width=.221\linewidth]{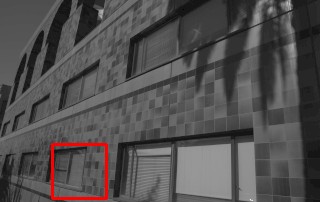} &
        \includegraphics[width=.14\linewidth]{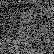} &
        \includegraphics[width=.14\linewidth]{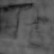} &
        \includegraphics[width=.14\linewidth]{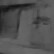} &
        \includegraphics[width=.14\linewidth]{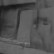} &
        \includegraphics[width=.14\linewidth]{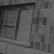} \\
        \includegraphics[width=.221\linewidth]{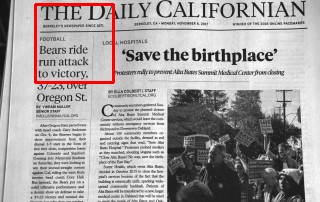} &
        \includegraphics[width=.14\linewidth]{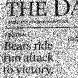} &
        \includegraphics[width=.14\linewidth]{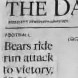} &
        \includegraphics[width=.14\linewidth]{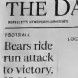} &
        \includegraphics[width=.14\linewidth]{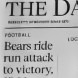} &
        \includegraphics[width=.14\linewidth]{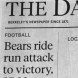} \\
        \includegraphics[width=.221\linewidth]{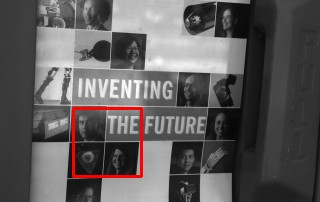} &
        \includegraphics[width=.14\linewidth]{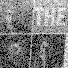} &
        \includegraphics[width=.14\linewidth]{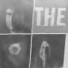} &
        \includegraphics[width=.14\linewidth]{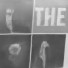} &
        \includegraphics[width=.14\linewidth]{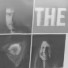} &
        \includegraphics[width=.14\linewidth]{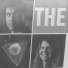} \\
        Full Image & Noisy & KPN \cite{MildenhallKPN18} & BPN \cite{BPN} & Ours & Ground Truth \\
    \end{tabular}
    \end{center}
    \caption{Burst denoising results on a KPN synthetic burst dataset \cite{MildenhallKPN18}. Our methods achieves better performances at reconstructing details such as texture and text.
    }
    \label{fig:vis_d}
\end{figure*}
\subsection{Evaluation Datasets}

Raw videos are captured with different sensors and different sensor gains, which lead to different noise strengths and different types of motions. The datasets we chose contain videos with such rich variations.

\vspace{4pt}
\noindent \textbf{KPN synthetic dataset} \cite{MildenhallKPN18} consists of 73 raw sequences with four different noise levels. Each noise level corresponds to a specific sensor gain. 
The relation between gains and the corresponding noise parameters ($\sigma_s$, $\sigma_s$) is listed as follows: $1\rightarrow$ $(2.7^{-3}, 6.8^{-3})$, $2\rightarrow$ $(6.2^{-3}, 1.5^{-2})$, $4\rightarrow$ $(1.4^{-2}, 3.6^{-2})$, $8\rightarrow$ $(3.3^{-2}, 8.3^{-2})$. 
Each sequence contains 8 grayscale raw frames. 
The misalignment is synthesized in the range $[2,16]$ pixels between 8 burst frames.

\vspace{4pt}
\noindent \textbf{CRVD Dataset} \cite{RViDeNet} consists of real raw videos of 12 scenes captured by a SONY IMX385 sensor. Videos in the first 6 scenes are taken as the training set and videos in the remaining 6 scenes are used as the evaluation set.
For each scene, five videos of 5 different ISOs ranging from 1,600 to 25,600 (corresponding gain from 1 to 16) are captured. The frames of each video only contain object motions without any camera motions. 
The ISOs correspond to the noise parameters ($(\sigma_s, \sigma_r)$): $1600\rightarrow(8.6^{-4}, 8.4^{-4})$, $3200\rightarrow(1.7^{-3}, 1.5^{-3})$, $6400\rightarrow(3.3^{-3}, 2.8^{-3})$, $12800\rightarrow(6.5^{-3}, 5.4^{-3})$, $25600\rightarrow(1.3^{-2}, 1.0^{-2})$.
We take clips of $N=3$ and $N=5$ frames as inputs to our method.

\vspace{4pt}
\noindent \textbf{HDR$+$ Dataset} \cite{hasinoff2016burst} consists of 3,640 bursts stored in DNG format \cite{DNG}.
The bursts are captured by a variety of Android mobile cameras (Nexus 5/6/5X/6P, Pixel, Pixel XL). The maximum number of burst frames is 10 and the maximum exposure time is 100ms.
The noise parameters can also be found in DNG format. Since the dataset cannot provide ground truth for quantitative evaluation, we perform qualitative evaluation on the bursts containing 8 frames captured in extreme low light scenes.

\begin{table*}[!t]
\centering
\footnotesize
\begin{tabular}{lcccc}
\toprule
& GFLOPs & Per-patch CPU runtime (ms) & Per-patch GPU runtime (ms) & Whole time (ms) 
\\
\midrule
BPN \cite{BPN} & 29.9 & 145.4  & 91.8 & 2754 \\
\hline
Ours &  14.3 & 72.6 & 44.5 & 1335 + 120 \\ \bottomrule
\end{tabular}
\caption{FLOPs and running times on gray-scale 1024 $\times$ 768 images (burst number $N=8$) on a Snapdragon 888 processor \cite{processor_888}. The images with $1024 \times 768$ are spilt into 48 128 $\times$ 128 patches.
We measure the running times of 128$\times$128 patch on CPU and GPU. According to their actual runing times, we assign different numbers of patches to CPU and GPU. Then we obtain the whole running time of denoising network. In our implementation, the additional running time for variance stabilization and alignment is 120 ms.}
\label{ab_runtimes}

\end{table*}

\subsection{Training}

We train our method on the CRVD dataset and KPN synthetic dataset.

\subsubsection{Training for KPN Dataset} \label{training_kpn}
At first, we perform unprocessing \cite{unprocessing} on sRGB images from the Open Images dataset \cite{oid} to obtain synthetic raw images.
The three channels in each sRGB image are averaged to produce a single-channel image. The single-channel image are transformed into the raw linear space. Then we synthesize motion \cite{MildenhallKPN18} on the single-channel images to simulate a clip of 8 frames. The mis-alignments between the reference frame and alternate frames are uniformly sampled in 2 to 16 pixels. 
When synthesizing raw noise, the sensor gain for each sequence is randomly sampled from $[1,4]$.
The Poisson-Gaussian noise are added to all frames according to the corresponding $\sigma_s, \sigma_r$. 
The loss function for training follows Eq. \eqref{loss_1}.
After adding noise, we perform multi-frame alignment on 8 frames of one burst. Then 8 aligned frames are taken as the network input.
All networks are adjusted to adapt the single-channel input.
The patch size is $256\times 256$ and the batch size is set to 16.
The learning rate is set as $10^{-4}$ for the first 50,000 iterations and $10^{-5}$ for the last 50000 iterations.

\setlength{\tabcolsep}{2pt}
\begin{figure*}[!t]
    \small
    \begin{center}
        \begin{tabular}{@{} c c @{}}
          \begin{tabular}{@{} c @{}}
               \includegraphics[width=.358\linewidth]{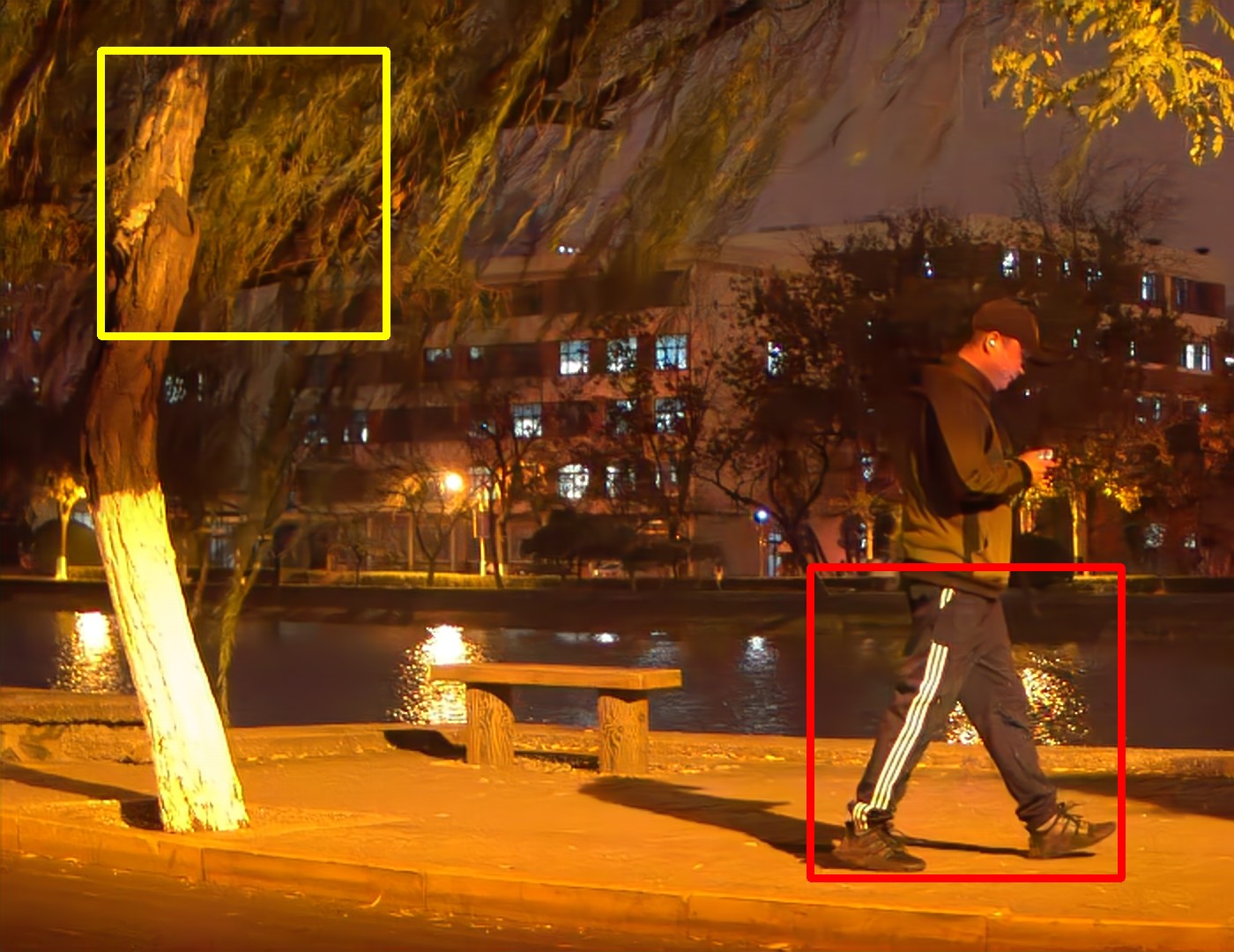} \\
               \includegraphics[width=.358\linewidth]{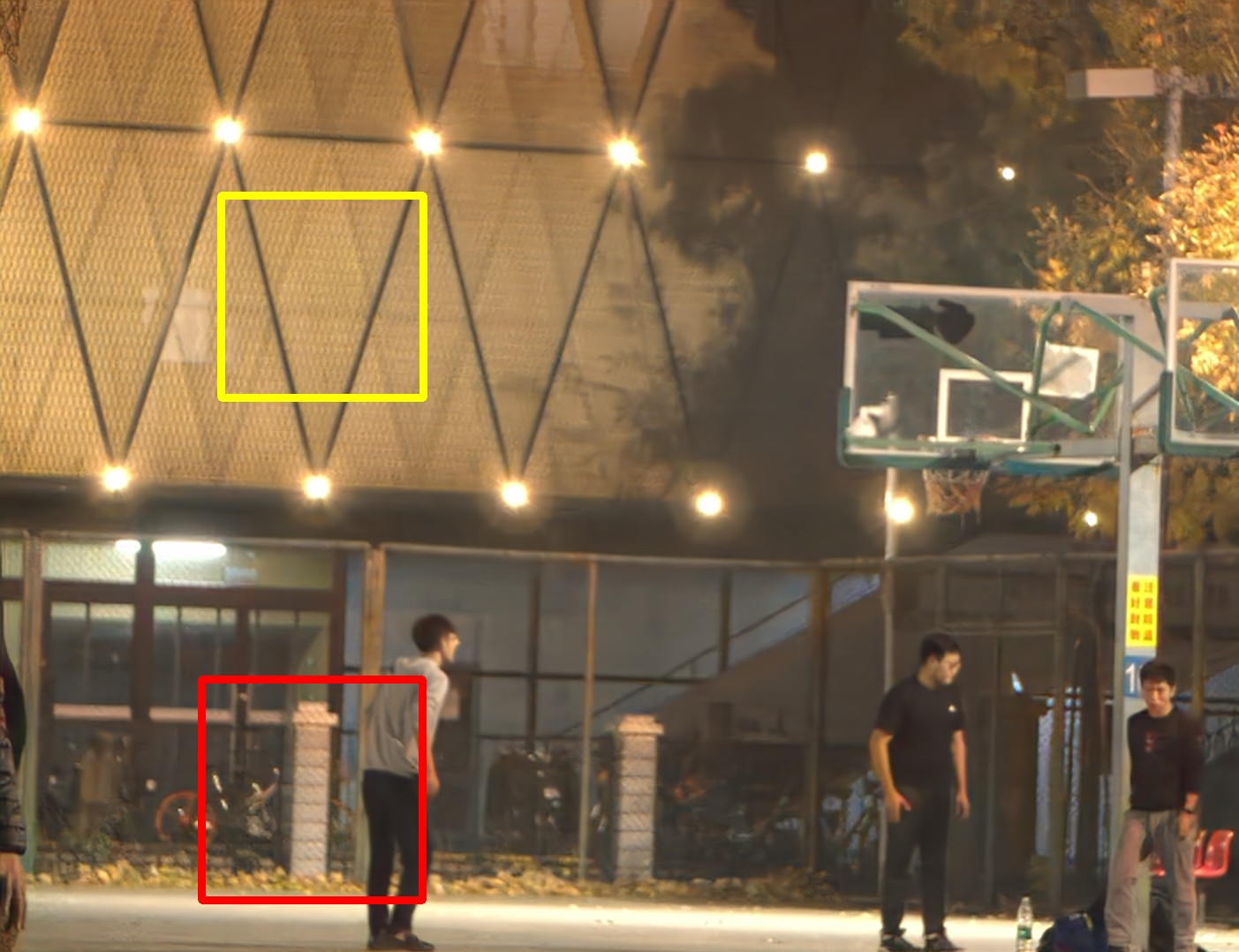} \\
               \includegraphics[width=.358\linewidth]{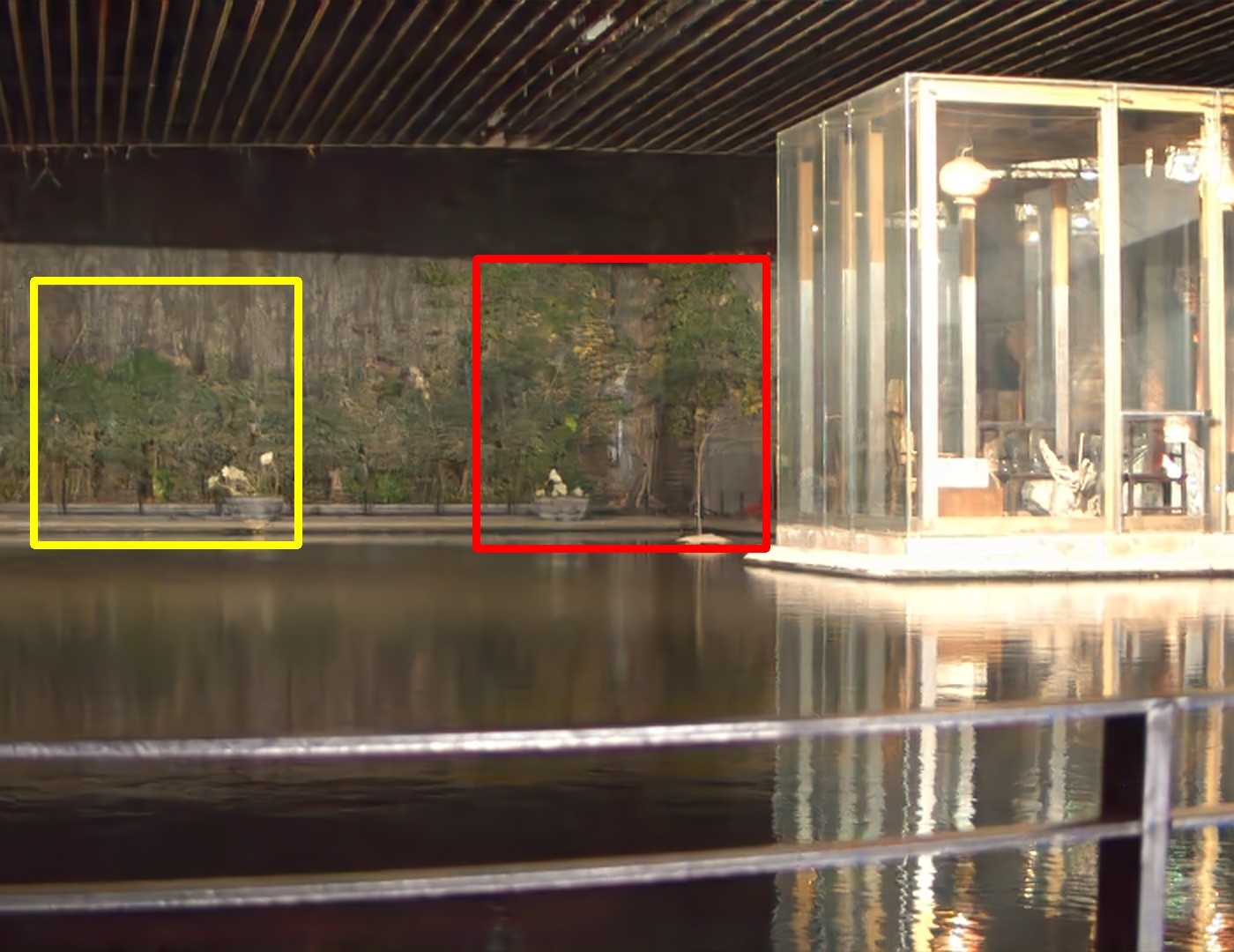} \\ 
               \includegraphics[width=.358\linewidth]{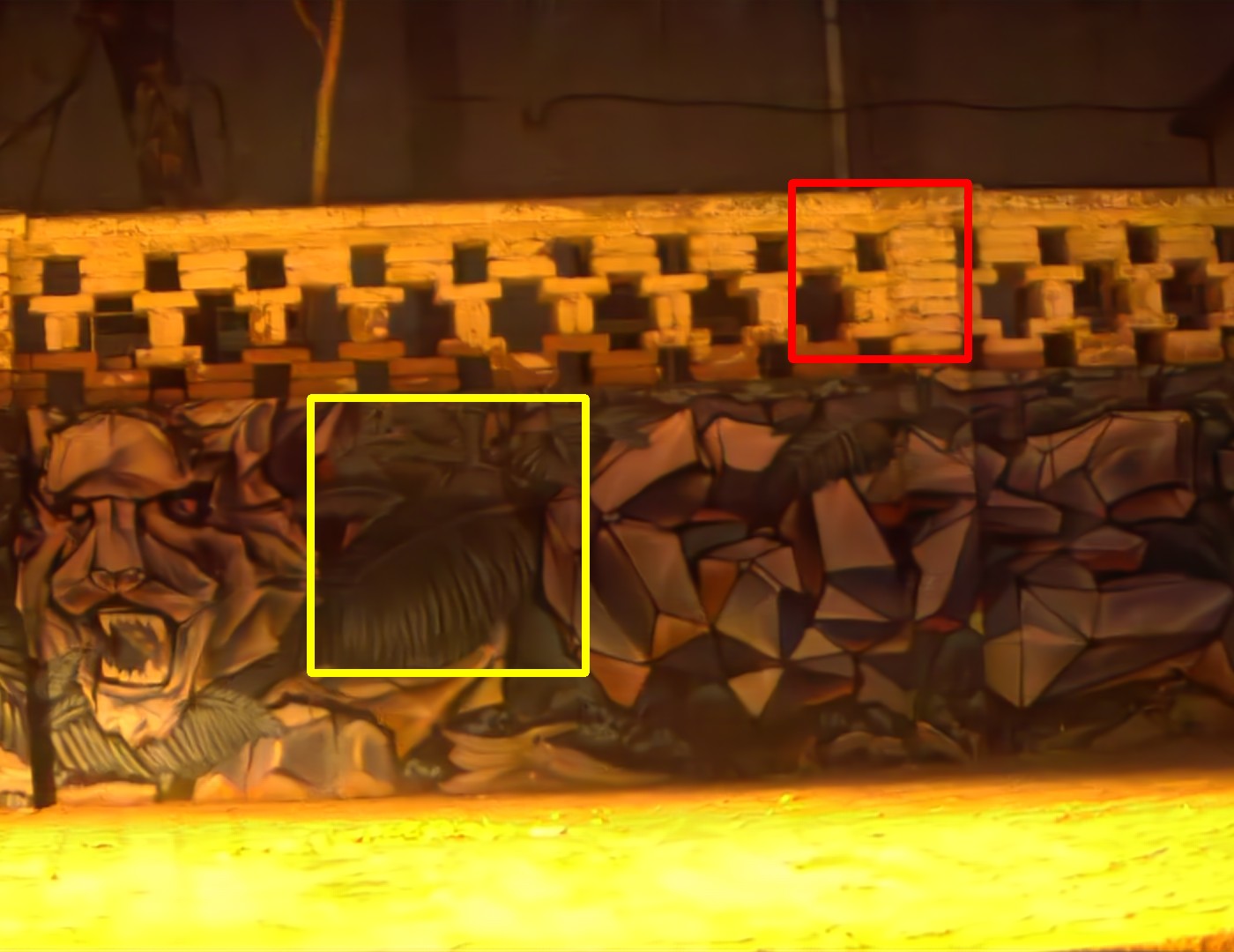} \\
               Full Image \\
          \end{tabular} & 
          \begin{tabular}{@{} c c c c @{}}
                \includegraphics[width=.135\linewidth]{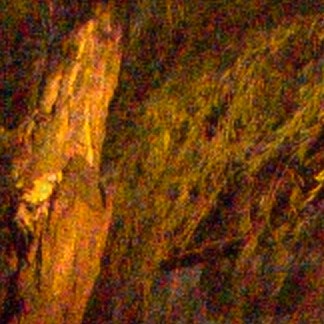} & 
                \includegraphics[width=.135\linewidth]{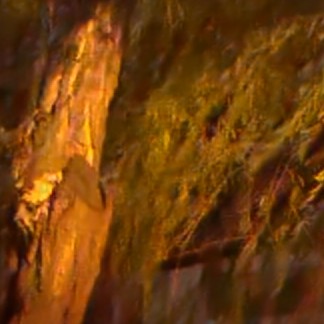} &
                \includegraphics[width=.135\linewidth]{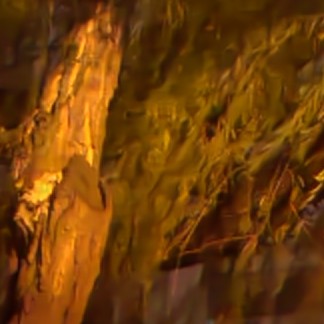} &
                \includegraphics[width=.135\linewidth]{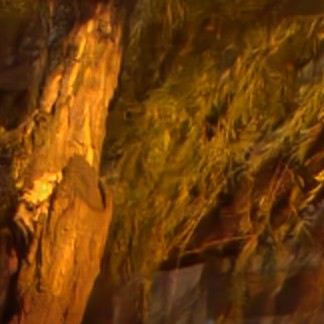} \\
                \includegraphics[width=.135\linewidth]{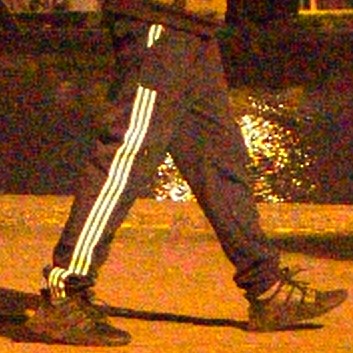} &
                \includegraphics[width=.135\linewidth]{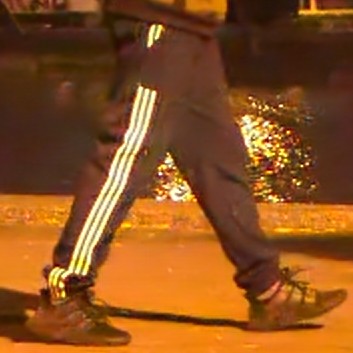} &
                \includegraphics[width=.135\linewidth]{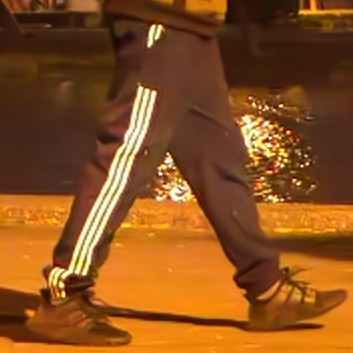} &
                \includegraphics[width=.135\linewidth]{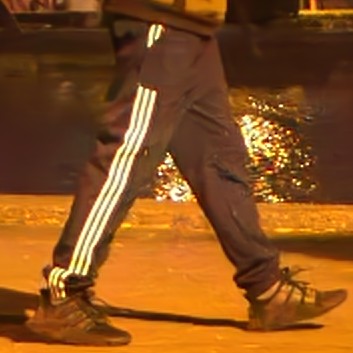} \\
                \includegraphics[width=.135\linewidth]{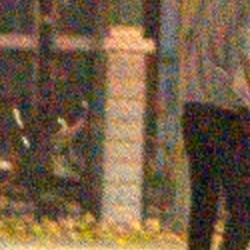} & 
                \includegraphics[width=.135\linewidth]{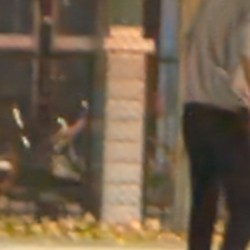} &
                \includegraphics[width=.135\linewidth]{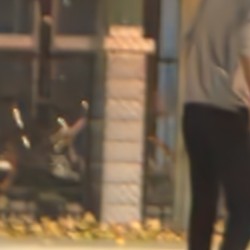} &
                \includegraphics[width=.135\linewidth]{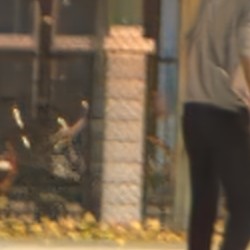} \\
                \includegraphics[width=.135\linewidth]{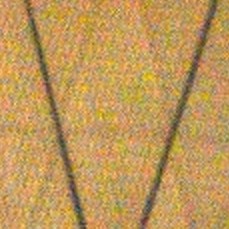} &
                \includegraphics[width=.135\linewidth]{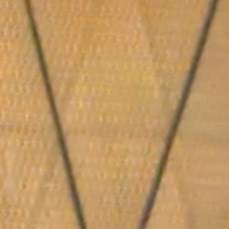} &
                \includegraphics[width=.135\linewidth]{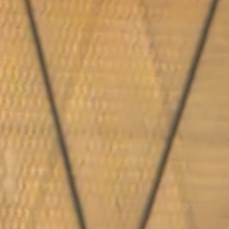} &
                \includegraphics[width=.135\linewidth]{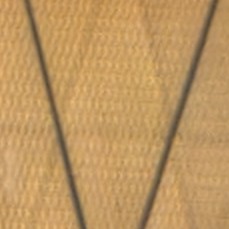} \\
                \includegraphics[width=.135\linewidth]{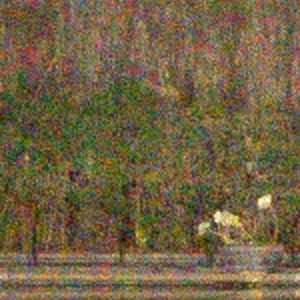} & 
                \includegraphics[width=.135\linewidth]{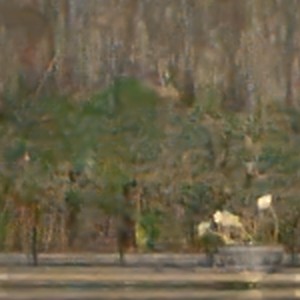} &
                \includegraphics[width=.135\linewidth]{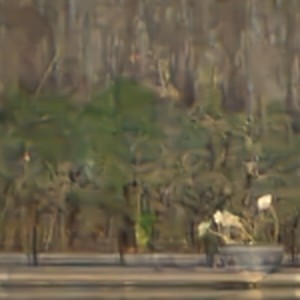} &
                \includegraphics[width=.135\linewidth]{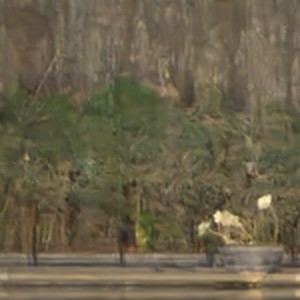} \\
                \includegraphics[width=.135\linewidth]{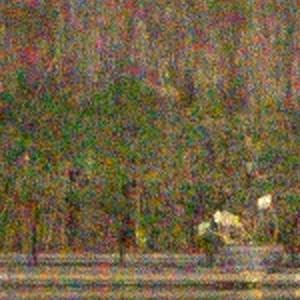} &
                \includegraphics[width=.135\linewidth]{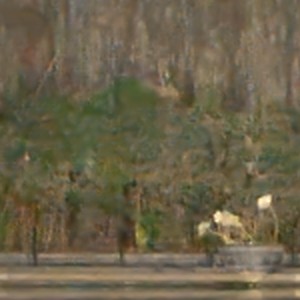} &
                \includegraphics[width=.135\linewidth]{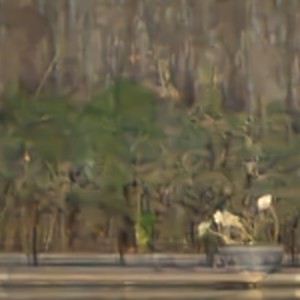} &
                \includegraphics[width=.135\linewidth]{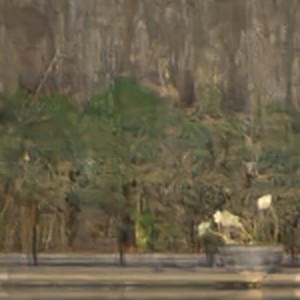} \\
                \includegraphics[width=.135\linewidth]{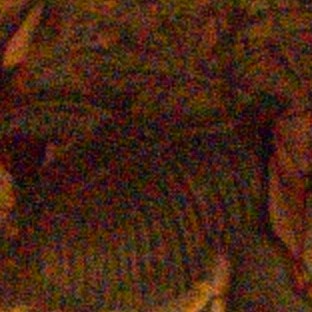} & 
                \includegraphics[width=.135\linewidth]{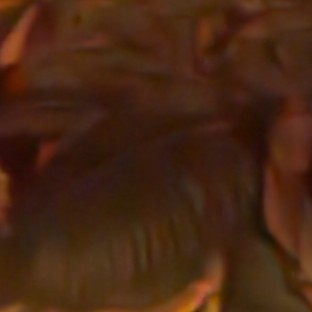} &
                \includegraphics[width=.135\linewidth]{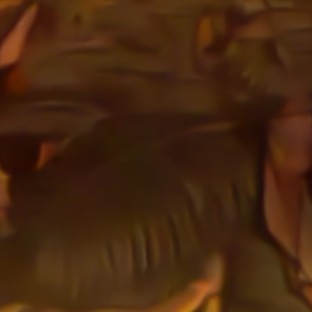} &
                \includegraphics[width=.135\linewidth]{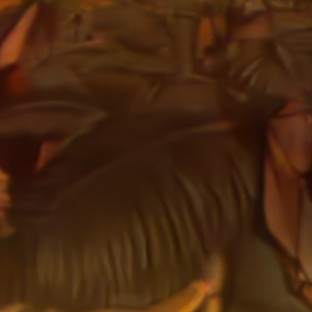} \\
                \includegraphics[width=.135\linewidth]{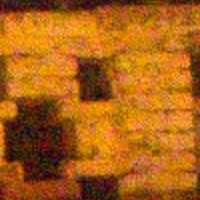} &
                \includegraphics[width=.135\linewidth]{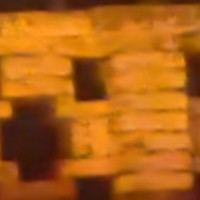} &
                \includegraphics[width=.135\linewidth]{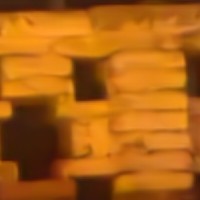} &
                \includegraphics[width=.135\linewidth]{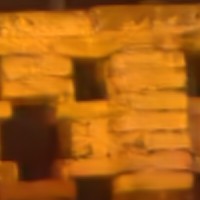} \\
                Noisy & FastDVD \cite{fastdvdnet} & RViDeNet \cite{RViDeNet} & Ours \\
          \end{tabular}
        \end{tabular}
    \end{center}
    \caption{Burst denoising results of different methods on real-world CRVD outdoor dataset \cite{RViDeNet}. Our methods recover more details of fine structures and moving objects.
    }
    \label{fig:crvd}
\end{figure*}
\setlength{\tabcolsep}{2pt}
\begin{figure*}[!t]
    \small
    \begin{center}
    \begin{tabular}{ccccc}
        \rotatebox{90}{\hspace{22pt}HDR+ \cite{hasinoff2016burst}} &
        \includegraphics[width=.26\linewidth]{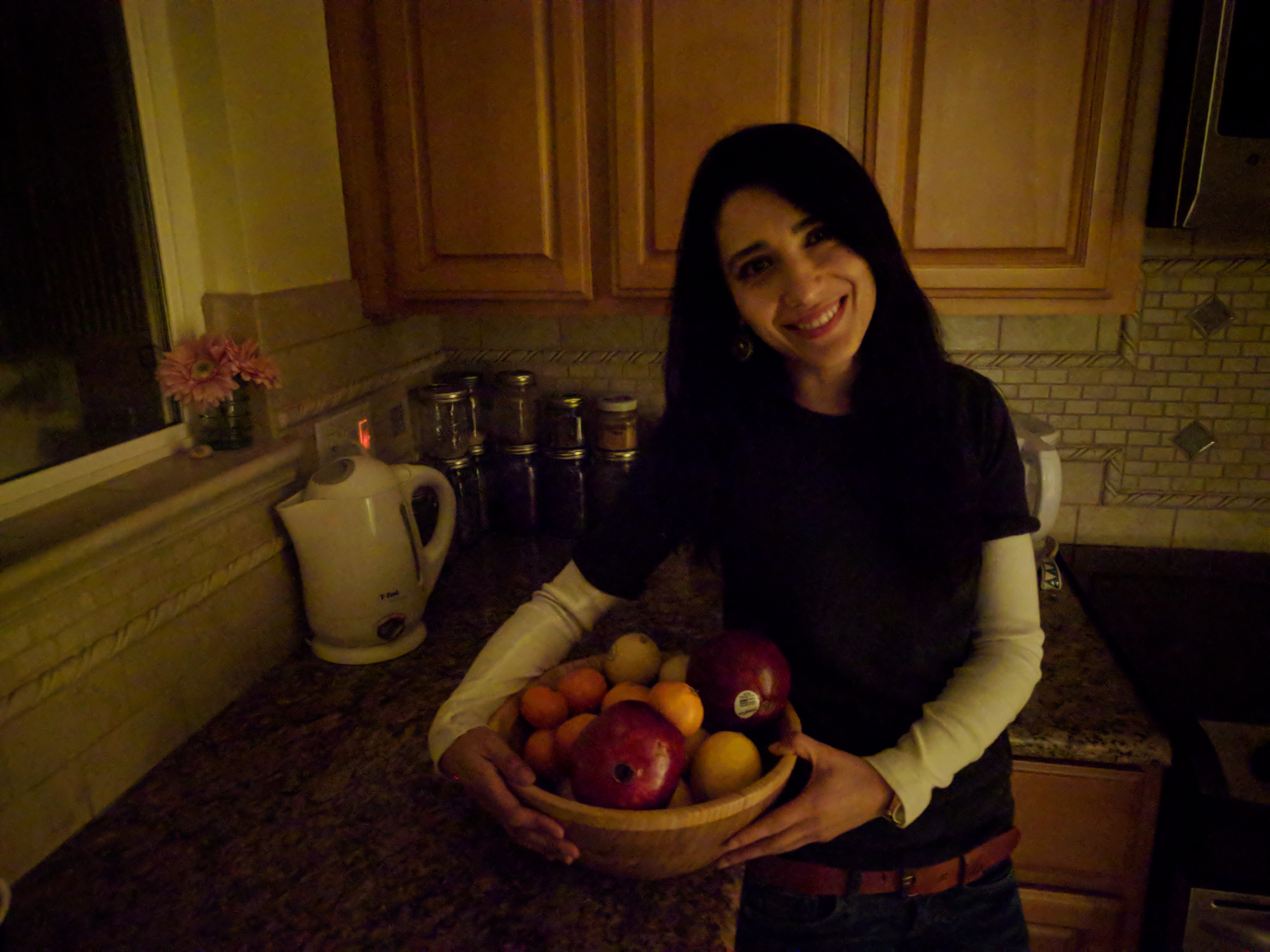} &
        \includegraphics[width=.2\linewidth]{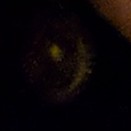} &
        \includegraphics[width=.2\linewidth]{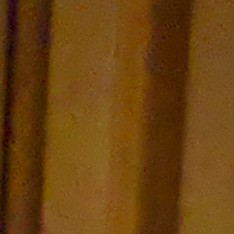} &
        \includegraphics[width=.2\linewidth]{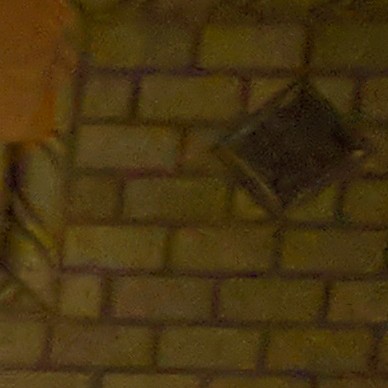} \\
        \rotatebox{90}{\hspace{22pt} VLL \cite{vll}} &
        \includegraphics[width=.26\linewidth]{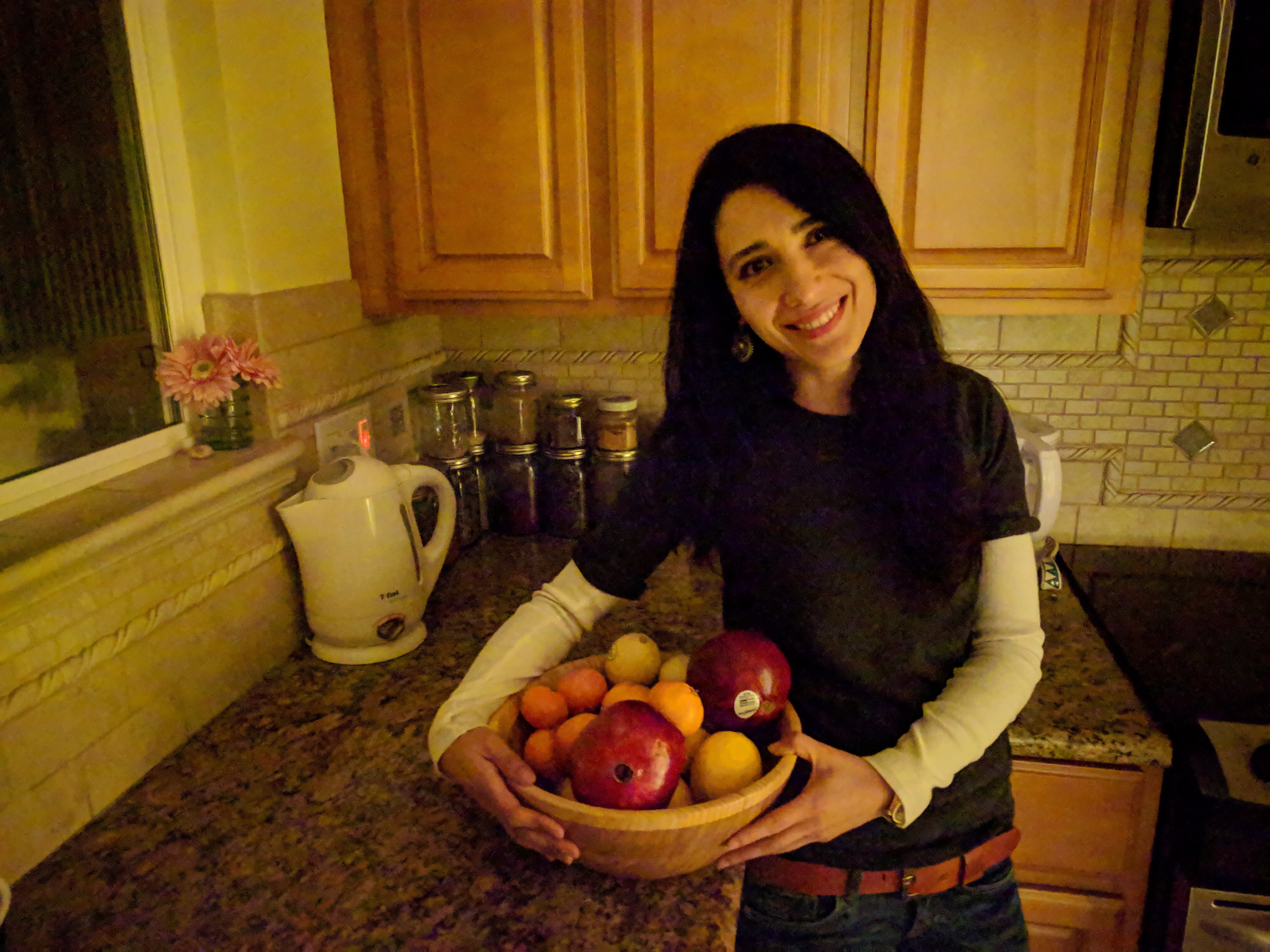} &
        \includegraphics[width=.2\linewidth]{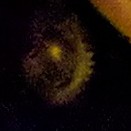} &
        \includegraphics[width=.2\linewidth]{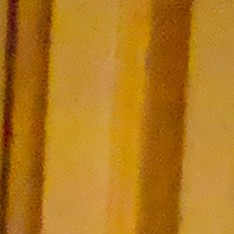} &
        \includegraphics[width=.2\linewidth]{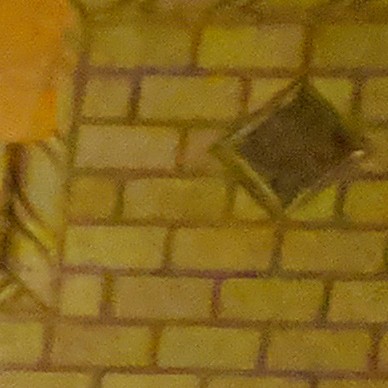} \\
        \rotatebox{90}{\hspace{30pt} Ours} &
        \includegraphics[width=.26\linewidth]{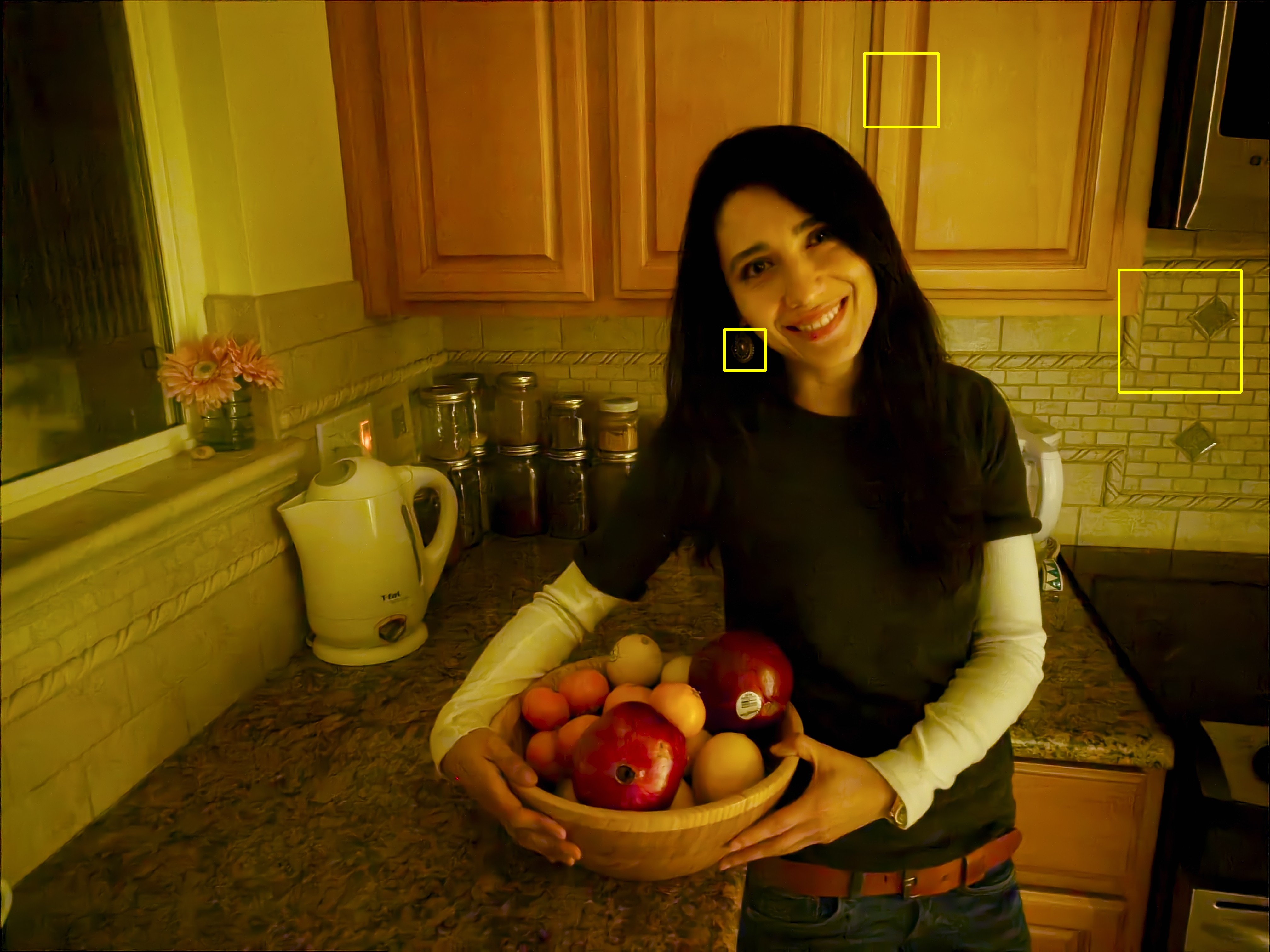} &
        \includegraphics[width=.2\linewidth]{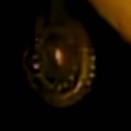} &
        \includegraphics[width=.2\linewidth]{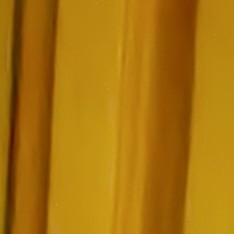} &
        \includegraphics[width=.2\linewidth]{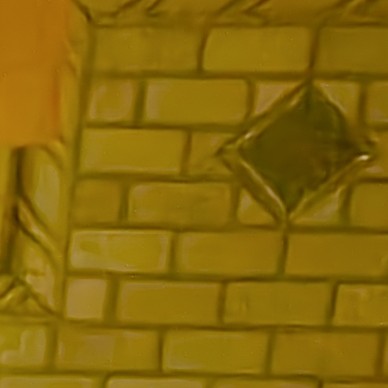} \\
        \rotatebox{90}{\hspace{22pt} HDR+ \cite{hasinoff2016burst}} &
        \includegraphics[width=.26\linewidth]{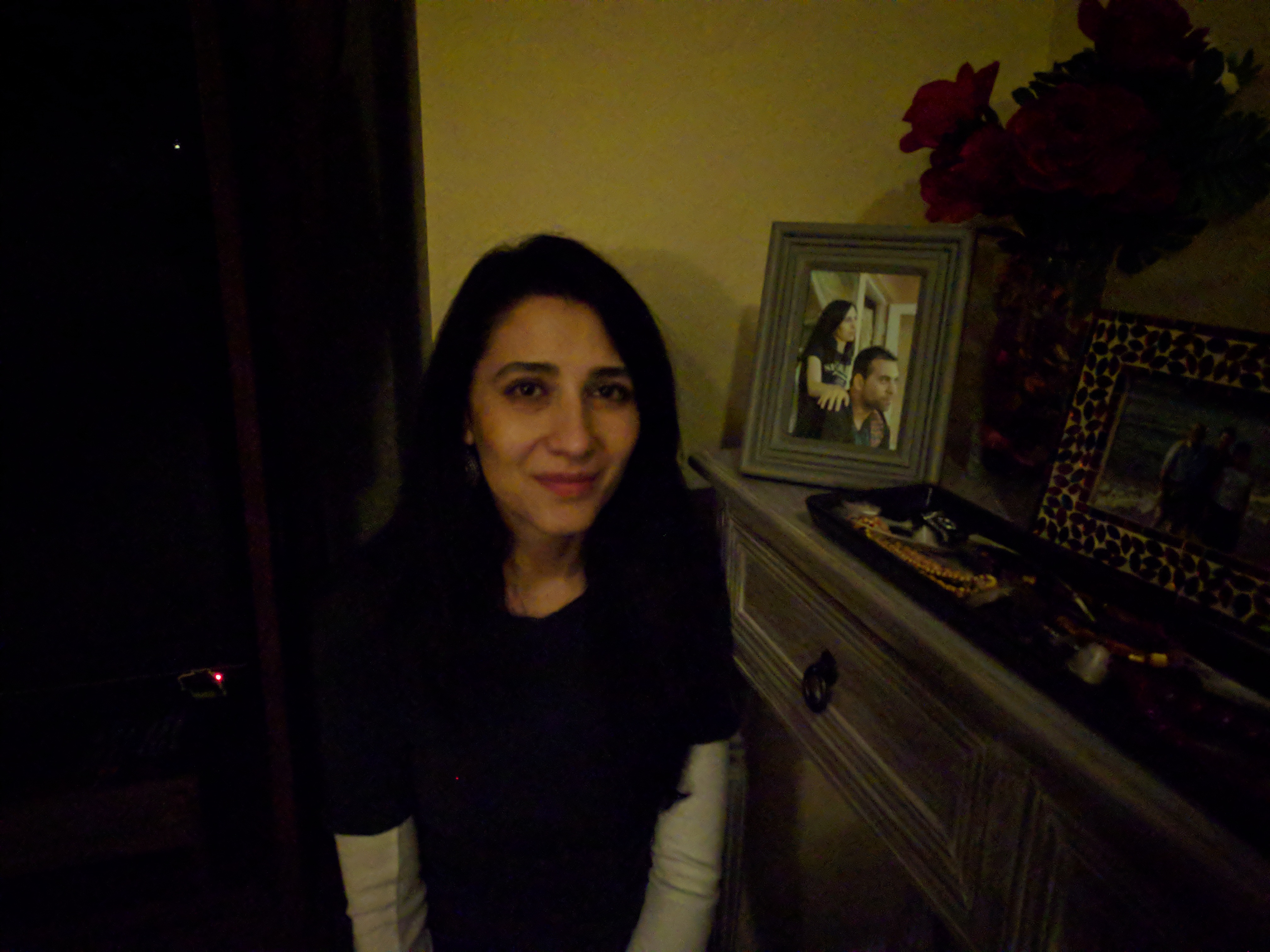} &
        \includegraphics[width=.2\linewidth]{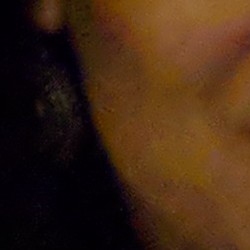} &
        \includegraphics[width=.2\linewidth]{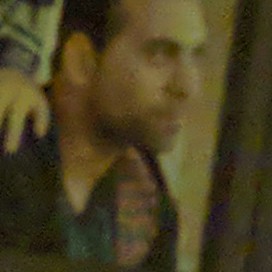} &
        \includegraphics[width=.2\linewidth]{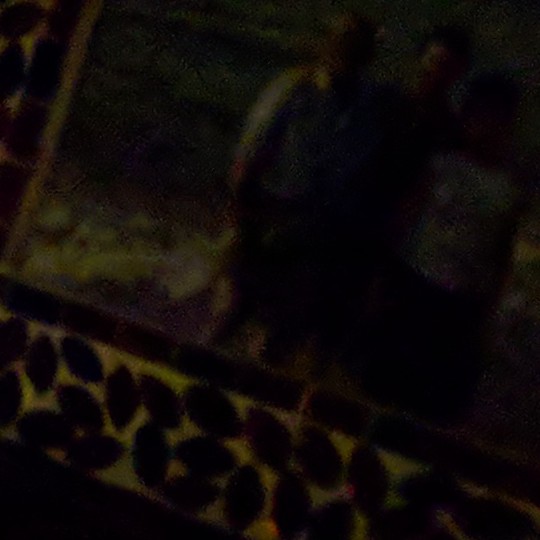} \\
        \rotatebox{90}{\hspace{22pt}VLL \cite{vll}} &
        \includegraphics[width=.26\linewidth]{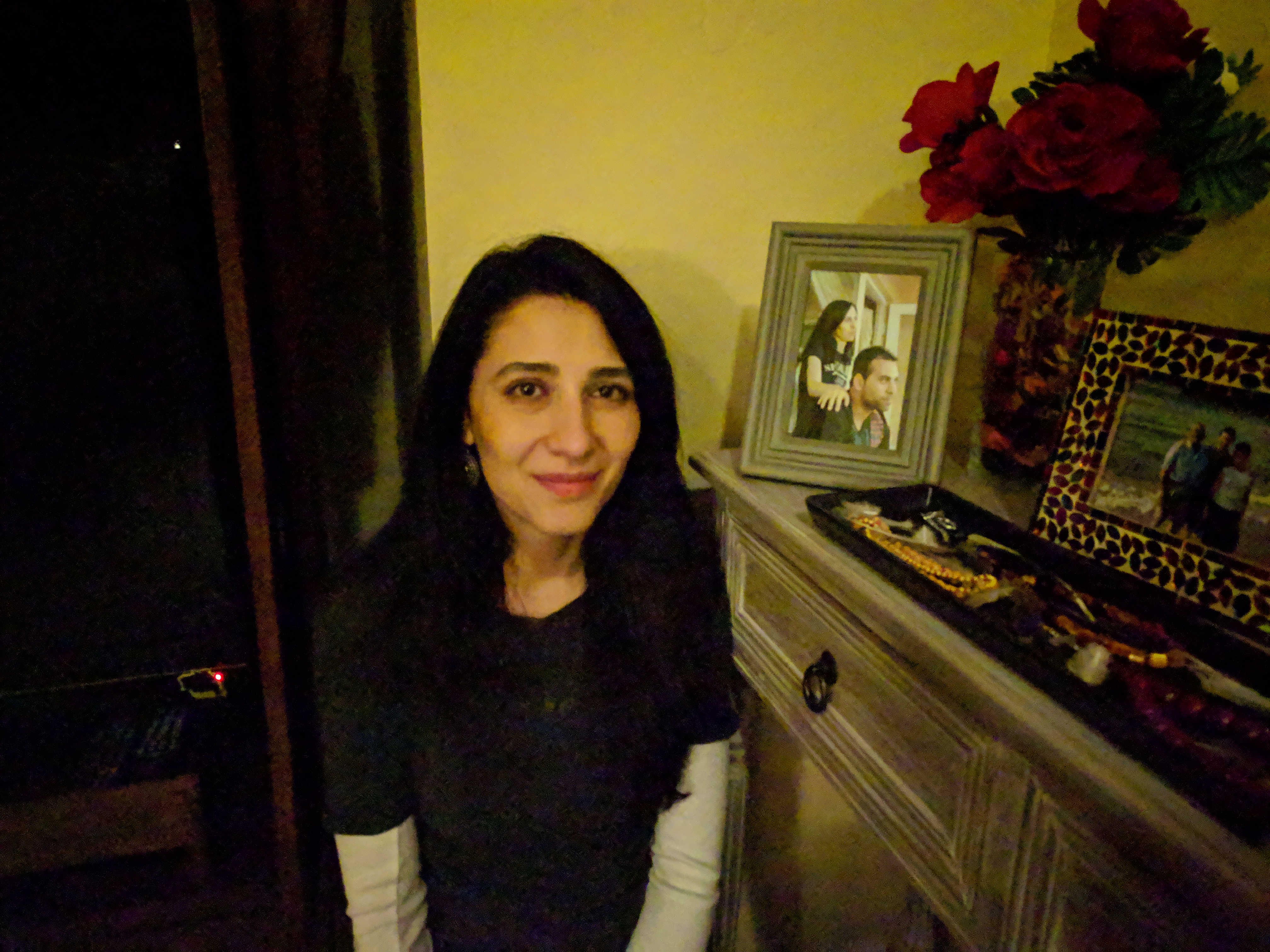} &
        \includegraphics[width=.2\linewidth]{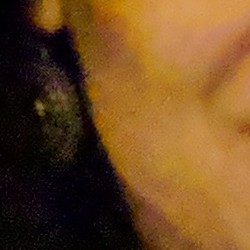} &
        \includegraphics[width=.2\linewidth]{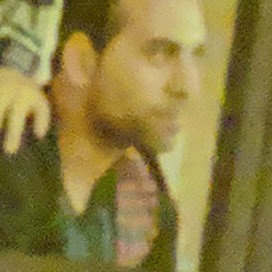} &
        \includegraphics[width=.2\linewidth]{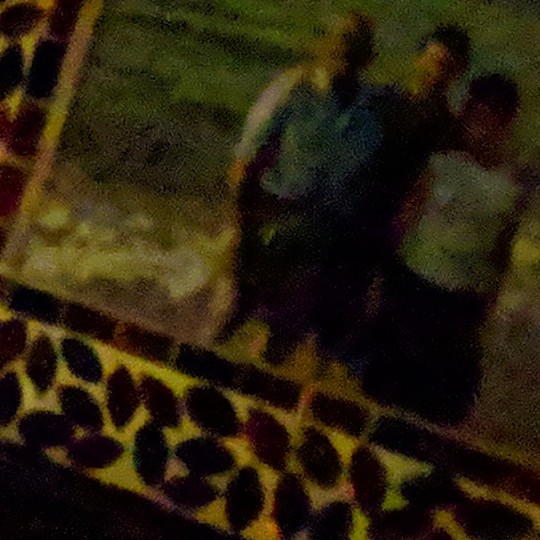} \\
        \rotatebox{90}{\hspace{30pt}Ours} &
        \includegraphics[width=.26\linewidth]{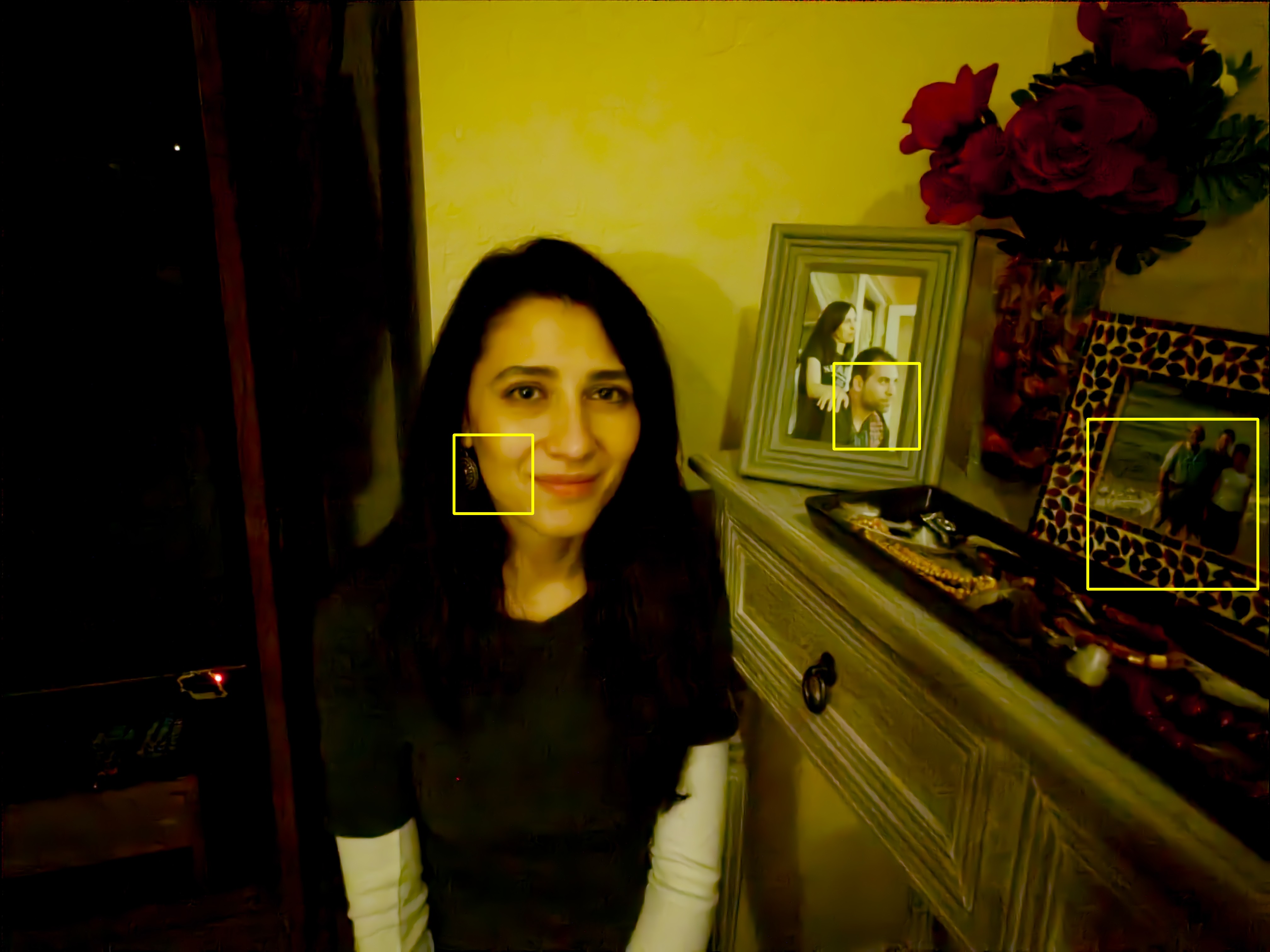} &
        \includegraphics[width=.2\linewidth]{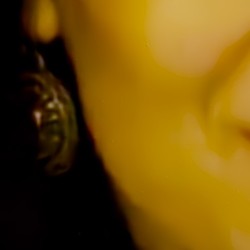} &
        \includegraphics[width=.2\linewidth]{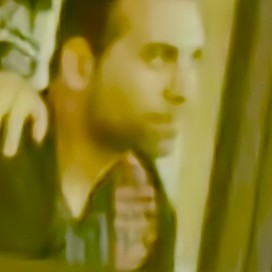} &
        \includegraphics[width=.2\linewidth]{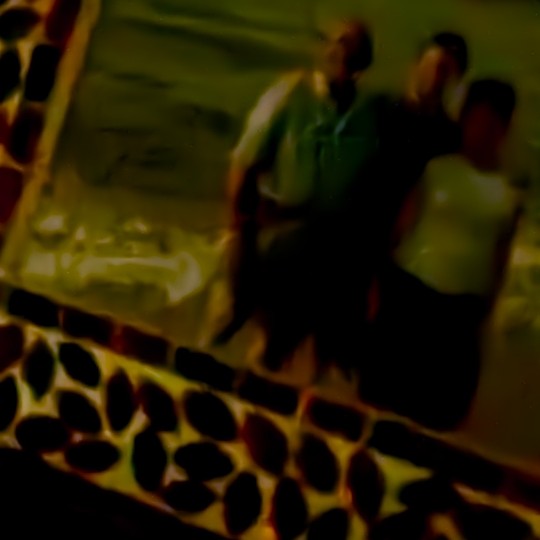} \\
    \end{tabular}
    \end{center}
    \caption{Burst denoising results on HDR$+$ dataset \cite{hasinoff2016burst}. Our method produces better images on extreme low light scenes with more details on edges and texture regions. 
    }
    \label{fig:hdrplus}
\end{figure*}

\subsubsection{Training for CRVD dataset}

Following the training setup of RViDeNet \cite{RViDeNet}, we train denoising models on SRVD dataset and finetune them on CRVD dataset (scene 1-6). 

\vspace{4pt}
\noindent\textbf{Training set creation.}
Since SRVD dataset is an sRGB video dataset, we perform the following operations to create the training pairs for raw video denoising.
To synthesize raw clean videos, three consecutive sRGB frames are randomly selected from SRVD videos and unprocessed into RGBG raw frames with the camera parameters (white balance, color correction matrix and Gamma factor provided in CRVD dataset). 
To construct training pairs, we add Poisson-Gaussian noise to synthesize three noisy frames from clean frames. The sensor gain is randomly sampled from $[1,16]$, which corresponds to ISO from 1,600 to 25,600. We can add Poisson-Gaussian noise to the frames according to the corresponding $\sigma_s, \sigma_r$.

For finetuning, CRVD dataset (scene 1-6) with raw videos is used, where we can obtain pairs of noisy input and clean ground truth. Clips of three consecutive noisy frames are randomly selected as input and ground truth are the clean frames corresponding to reference noisy frames.

\vspace{3pt}
\noindent\textbf{Training settings.}
For our proposed method, we stabilize and align the three noisy frames before feeding them into the proposed denoising network. Then the multi-frame denoising network produces the clean outputs in the variance-stabilization space, which are then transformed back into the raw linear space.
The loss function for training adopts Eq. \eqref{loss_1}. The loss function of finetuning adopts Eq. \eqref{eq:loss_2}.
All the networks are trained with learning rate $10^{-4}$ for 85,000 iterations and finetuned with learning rate $10^{-5}$ for 30,000 iterations.
The proposed network is implemented in PyTorch \cite{NEURIPS2019_9015} and trained with NVIDIA 1080TI GPUs.

\subsection{Evaluation}
We evaluate our method and compare it with state-of-the-art multi-frame denoising methods, including V-BM4d \cite{DBLP:conf/ipas/MaggioniBFE11}, FastDVDNet \cite{fastdvdnet}, RViDeNet \cite{RViDeNet}, KPN \cite{MildenhallKPN18} and BPN \cite{BPN}, for evaluation on the two datasets. The compared methods are adopted from authors' original implementations.

\subsubsection{KPN Synthetic Dataset}

Table \ref{result_kpn} reports the results on KPN grayscale test set \cite{MildenhallKPN18}. The PSNR and SSIM are computed after gamma correction to reflect perceptual quality.
As for BPN \cite{BPN}, we directly utilize its released pretraind model for evaluation. Since we cannot access the original models of KPN \cite{MildenhallKPN18}, we train KPN model based on the implementation accepted by the original authors.
Their performances are shown on Table \ref{result_kpn}.
As for our method, we set the group number $k=3$. Then 7 alternate frames are divided into 3 groups. 8 frames will be processed by 4 efficient multi-frequency networks sequentially.
Our method shows great improvements about at all levels over KPN \citep{MildenhallKPN18} and BPN \cite{BPN}. As for extreme noisy case (Gain $\propto$ 8), we improve 0.72 PSNR against \cite{BPN}.

\subsubsection{CRVD Dataset}
We train all the methods in raw linear space. Then results in the raw domain are further processed into the sRGB domain by the pretrained-ISP model described in RViDeNet. Then PSNRs and SSIMs are calculated in the sRGB domain.
For the evaluation of $N=5$, we train RViDeNet \cite{RViDeNet} and FastDVDNet \cite{fastdvdnet} based on their implementation with the same settings as ours. 

Table \ref{table_crvd} lists the average PSNR and SSIM of raw domain and sRGB domain for video scenes 6-11. 
When we use only $N=3$ frames for denoising, it can be observed that our method outperforms the compared denoising methods. Compared with the state-of-the-art RViDeNet \cite{RViDeNet}, our improvement is 0.35dB PSNR in raw domain and 0.46dB PSNR in sRGB domain.
When the network takes $N=5$ frames as inputs, our methods still achieves the best performance. 
We visualize the denoising results of outdoor scenes in Figure~\ref{fig:crvd}.
The proposed model generates better pleasing details in both static and dynamic regions. 

\setlength{\tabcolsep}{0.5pt}
\begin{figure}[!t]
    \small
    \begin{center}
    \begin{tabular}{cccc}
        \includegraphics[width=.245\linewidth]{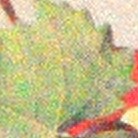} &
        \includegraphics[width=.245\linewidth]{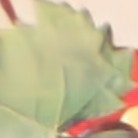} &
        \includegraphics[width=.245\linewidth]{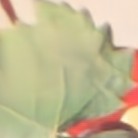} &
        \includegraphics[width=.245\linewidth]{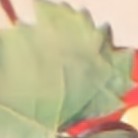} \\
        Noisy & w/o integration & noise map \cite{MildenhallKPN18} & $k-$sigma \cite{PMRID} \\
        \includegraphics[width=.245\linewidth]{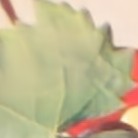} &
        \includegraphics[width=.245\linewidth]{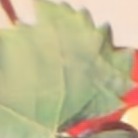} &
        \includegraphics[width=.245\linewidth]{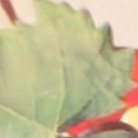} &
        \includegraphics[width=.245\linewidth]{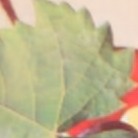} \\
        VS-gain & GAT \cite{GAT} & Ours & Ground Truth \\
        \includegraphics[width=.245\linewidth]{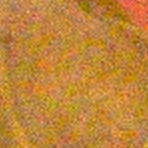} &
        \includegraphics[width=.245\linewidth]{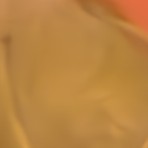} &
        \includegraphics[width=.245\linewidth]{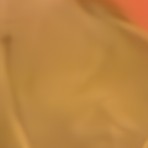} &
        \includegraphics[width=.245\linewidth]{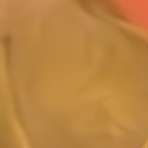} \\
        Noisy & w/o integration & noise map \cite{MildenhallKPN18} & $k-$sigma \cite{PMRID} \\
        \includegraphics[width=.245\linewidth]{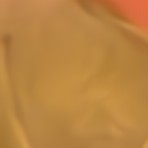} &
        \includegraphics[width=.245\linewidth]{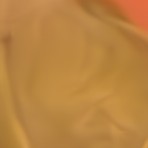} &
        \includegraphics[width=.245\linewidth]{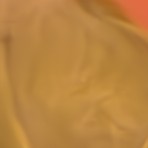} &
        \includegraphics[width=.245\linewidth]{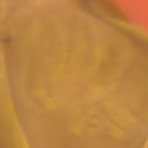} \\
        VS-gain & GAT \cite{GAT} & Ours & Ground Truth \\
        \includegraphics[width=.245\linewidth]{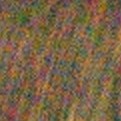} &
        \includegraphics[width=.245\linewidth]{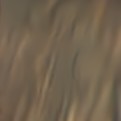} &
        \includegraphics[width=.245\linewidth]{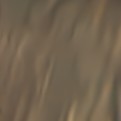} &
        \includegraphics[width=.245\linewidth]{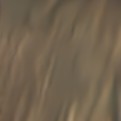} \\
        Noisy & w/o integration & noise map \cite{MildenhallKPN18} & $k-$sigma \cite{PMRID} \\
        \includegraphics[width=.245\linewidth]{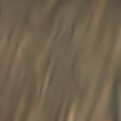} &
        \includegraphics[width=.245\linewidth]{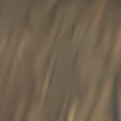} &
        \includegraphics[width=.245\linewidth]{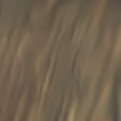} &
        \includegraphics[width=.245\linewidth]{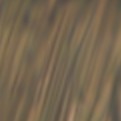} \\
        VS-gain & GAT \cite{GAT} & Ours & Ground Truth \\
    \end{tabular}
    \end{center}
    \caption{Visual comparisons of different noise prior integration on CRVD dataset \cite{RViDeNet} (burst number $N=3$).
    }
    \label{fig:vis_noise_prior}
\end{figure}
\subsubsection{Qualitative Evaluation on HDR$+$ Dataset}
For evaluation on HDR$+$ dataset \cite{hasinoff2016burst}, we train the color version of our denoising network. The training settings are similar to training for KPN synthetic dataset as described in Section \ref{training_kpn}. 
We compare our method with HDR$+$ \cite{hasinoff2016burst} and VLL \cite{vll} as shown Fig.~\ref{fig:hdrplus}. These two images (from Nexus 6p) are captured with ISO 5760 and 100ms exposure time.
Since the post-processing pipeline of HDR$+$ \cite{hasinoff2016burst} is not released to the public, we utilize the post-processing pipeline in RViDeNet \citep{RViDeNet} to transform raw outputs into the sRGB space. Therefore, there exists inevitable color shift between our method and HDR$+$. Our method generally produces less noisy than compared methods. More details on edges and texture regions are recovered by our methods.

\subsection{Computational Expense}
We first report the number of floating point operations (FLOPs) required in our denoising network and BPN \cite{BPN} in table \ref{ab_runtimes}. It is observed that our denoising network requires fewer FLOPS than BPN.
Then we evaluate our method and BPN \cite{BPN} on on the smartphone with the Snapdragon 888 processor \cite{processor_888}.
Snapdragon 888 processor contains one Qualcomm Kryo CPU and one Qualcomm Adreno 660 GPU. We implement the variance stabilization and alignment on 
CPU processor. These two algorithms are accelerated by ARM Neon Intrinsics \cite{Arm}. 
As for 8 images with resolution 1024 $\times$ 768, running time of variance stabilization is 20 ms and running time of alignment is 100 ms. 
Then we utilize the PPLNN platform \cite{PPLNN} to measure the running time of denoising networks on Snapdragon 888 processor.
Because of the limited memory on smartphones, 1024 $\times$ 768 images are divided into 48 128 $\times$ 128 non-overlaping patches, which are processed separately.
To accelerate execution on smartphones, we run our denoising network and BPN on both CPU and GPU processors. 
The per-patch and overall running times on CPU and GPU are listed in Table \ref{ab_runtimes}. We assign different numbers of patches on CPU and GPU processors according to their actual running times. Taking our denoising network as an example, we assign 30 patches to GPU processor and 18 patches to CPU processor, which leads to the most efficient running time of denoising network ($30 \times 44.5\text{ms} = 1335$ms).

\begin{table}[!t]
\centering
\footnotesize
\begin{tabular}{lcc}
\toprule
Algorithm  & raw & sRGB \\ 
\midrule
Ours w/o integration &   43.99 / 0.988 &  39.70 / 0.979 \\
Ours w/ noise map \cite{MildenhallKPN18}  & 44.26 / 0.988 & 40.16 / 0.980  \\ 
Ours w/ $k-$sigma \cite{PMRID}  & 44.28 / 0.988  & 40.18 / 0.980 \\
Ours w/ VS-gain  & 44.29 / 0.988 & 40.19 / 0.980 \\ 
Ours w/ GAT \cite{GAT}  & 44.34 / 0.989 & 40.44 / 0.981 \\ \hline
Ours    & 44.43 / 0.989  & 40.51 / 0.982 \\ \bottomrule
\end{tabular}
\caption{Ablation study of our method on noise prior integration evaluated on CRVD dataset (burst number $N=3$). }
\label{ab_noiseprior}
\end{table}

\begin{table*}[!t]
\centering
\footnotesize
\begin{tabular}{lccccc}
\toprule
Algorithm & Gain $\propto$ 1 & Gain $\propto$ 2 & Gain $\propto$ 4 & Gain $\propto$ 8 & Average \\
\midrule
KPN \cite{MildenhallKPN18}  & 36.47 &  33.93 & 31.19  & 27.97 & 32.39  \\ 
KPN* \cite{MildenhallKPN18}  & 36.47 &  33.94 & 31.21 & 28.24 & 32.47 \\ 
KPN* + alignment & 37.14 &  34.62 & 31.87  & 28.73 & 33.09 \\ 
BPN \cite{BPN}  & 38.18 &  35.42 & 32.54  & 29.45 & 33.90  \\ 
BPN + alignment & 38.28 &  35.79 & 32.96  & 29.86 & 34.22 \\
\hline
Ours w/o alignment & 37.63 & 34.92 & 32.13 & 28.89 & 33.39 \\
Ours $+$ alignment & 38.48 &  35.79 & 32.87  & 29.62 & 34.19 \\
Ours & 39.39 &  36.52 & 33.47  & 30.20 &  34.90 \\ 
\bottomrule
\end{tabular}
\caption{Ablation study of influence of alignment on KPN dataset ((burst number $N=8$)).}
\label{ab_align_1}

\begin{tabular}{lccc}
\toprule
\setlength\tabcolsep{10pt}
Algorithm   &   CRVD & $\pm16$ CRVD & running time(ms) \\
\midrule
Ours w/o alignment  & 44.36 / 0.989  &  43.68 / 0.978  & - \\ 
Ours w/ DCN \cite{RViDeNet,wang2019edvr} &  44.46 / 0.989 & 44.11 / 0.981 & 2957 \\ 
Ours  & 44.43 / 0.989 & 44.05 / 0.980  & 38 \\ \bottomrule
\bottomrule
\end{tabular}
\caption{Ablation study of influence of different alignment methods on CRVD dataset \cite{RViDeNet} (burst number $N=3$). ``$\pm16$ CRVD'' denotes performing $\pm16$ pixels misalignment on CRVD dataset. Running times of the alignment modules is calculated on the Snapdragon 888 mobile processor.}
\label{ab_align}
\end{table*}

\subsection{Ablation Study}\label{sec:ablation}

We conduct ablation study to demonstrate the effectiveness of \textit{noise prior integration}, \textit{multi-frame alignment} and \textit{multi-frame denoising}.
These improvements are evaluated to illustrate that our methods run with limited computational cost but with competitive performance.

\begin{table}[t]
\centering
\footnotesize
\setlength\tabcolsep{3pt}
\begin{tabular}{lcccc}
\toprule
Algorithm & Raw & sRGB \\ \midrule
Ours-SQ1 & 44.37 / 0.989 & 40.52 / 0.982 \\  
Ours-SQ2 & 44.46 / 0.989 & 40.56 / 0.982 \\ 
Ours-SQ3 & 44.61 / 0.990 & 40.76 / 0.983 \\ \hline  
Ours & 44.70 / 0.990  &  40.88 / 0.983 \\ \bottomrule
\end{tabular}
\caption{Comparison of different sequential denoising strategies on CRVD dataset (burst number $N=5$).}
\label{ab_seq_denoising}
\end{table}
\begin{table}[t]
\centering
\footnotesize
\setlength\tabcolsep{3pt}
\begin{tabular}{lcc}
\toprule
Algorithm & Raw & sRGB \\
\midrule
Ours w/o aggregation & 44.17 / 0.988 & 40.10 / 0.980 \\  
Ours w/ pointwise conv & 43.71 / 0.987 & 39.69 / 0.979 \\ 
Ours w/ attention \cite{disneyasy} & 44.28 / 0.988 & 40.34 / 0.981 \\ \hline
Ours   & 44.43 / 0.989  &  40.51 / 0.982 \\ \bottomrule
\end{tabular}
\caption{Ablation study of multi-frequency aggregation on CRVD dataset (burst number $N=3$).}
\label{ab_denoising_network}
\end{table}

\vspace{3pt}
\noindent\textbf{Noise prior integration.}
Since CRVD dataset \cite{RViDeNet} is the only multi-frame raw dataset with ground truth, we perform ablation study of noise prior integration on CRVD dataset (burst number $N=3$).
In Table \ref{ab_noiseprior}, we evaluate the different methods for implementing \textit{noise prior integration} on CRVD dataset. 
We first remove the noise prior integration stage (denoted as ``Ours w/o prior integration''), to let the network to handle unstable variances directly. Removing noise prior integration leads to about 0.43dB drop. 
Then we compare our method with the mainstreaming noise adaption method: noise map \cite{MildenhallKPN18, fastdvdnet, EMVD}.
In training and finetuning, we replace the noisy input of network in our system by concatenation of noise map and noisy images (denoted as ``Ours w/ noise map''). This method brings a drop of 0.17dB PSNR. 

To further analyze the importance of variance stabilization, we evaluate the effect of stabilizing sensor gains and brightness in Poisson distribution. we first investigate other techniques of stabilizing sensor gains. 
We experiment using Eq. \eqref{eq:acute_x} (denoted as ``Ours w/ VS-gain'') and $k-$sigma transform Eq. \eqref{eq:k_sigma} proposed in PMRID \cite{PMRID}  (denoted as ``Ours w/ $k-$sigma''). 
Surprisingly, they achieves almost the same performances and shows a drop of about 0.14dB PSNR compared with our method. It reveals that $k-$sigma transform \cite{PMRID} can also be used to stabilize sensor gains.
Finally, we test stabilizing the variances of different brightness in Poisson distribution. We compare our generalized Tukey-Freeman transformation with well-known generalized anscombe transformation (GAT) \cite{GAT} ((denoted as ``Ours w/ GAT'')). Using GAT brings a drop of 0.09dB PSNR. For the reason that our generalized Tukey-Freeman transformation surpasses generalized anscombe transformation in raw denoising, please refer to Appendix~\ref{secA1}.

Fig~\ref{fig:vis_noise_prior} shows the visualization of denoising results with different types of noise prior integration. Our method and GAT \cite{GAT} show more details and textures than noise map \cite{MildenhallKPN18, zhang2018ffdnet} and $k-$sigma transformation \cite{PMRID}. Our method also achieves slight improvements on recovering edges against GAT \cite{GAT}.

\vspace{3pt}
\noindent\textbf{Multi-frame alignment.}
  We demonstrate that performing explicit alignment is necessary in our denoising system and state-of-the-art burst denoising methods. As the mainstream burst denoising methods, kernel prediction methods \cite{MildenhallKPN18, BPN} do not require an explicit alignment module. In contrast, RViDeNet \cite{RViDeNet} and BDNet \cite{BDNet} utilize Deformable Convolution \cite{DCN,DCN2} as an explicit alignment module before multi-frame denoising. To demonstrate the effectiveness of explicit alignment, we integrate our alignment module into two kernel prediction methods: KPN \cite{MildenhallKPN18} and BPN \cite{BPN}. 
  
 As there is no KPN model \cite{MildenhallKPN18} released to the public, we implement KPN and report its result (denoted as ``KPN*'' in Table \ref{ab_align_1}). Our implementation shows similar performance compared with the original results in \cite{MildenhallKPN18}. Adding an alignment module into KPN* (denoted as ``KPN* + alignment'') leads to about 0.7dB at gain $\propto$ 4. The results in Table \ref{ab_align_1} also show that adding our alignment module on top of BPN \cite{BPN} achieves a increase of about 0.4 dB PSNR at gain $\propto$ 2, 4, 8. When we remove the alignment module ( denoted as ``Ours w/o alignment'') in our method, our denoising network is trained on the synthetic frames with misalignment in $[2,16]$ pixels. It achieves the approximate performance with ``KPN* + alignment''.Then we perform our alignment (denoted as ``Ours + alignment'') on the above model (trained in ``Ours w/o alignment''). It is shown that directly applying our alignment brings improvement of 0.84dB average PSNR. Finally, we train our denoising network on images aligned by our method (denoted as ``Ours''). The result shows that training on aligned images brings about an increase of 0.72 dB average PSNR.

  In Table \ref{ab_align}, we compare our alignment with learning-based alignment in CRVD dataset. 
  Similar to KPN synthetic dataset, we synthesize large motion (up to $\pm 16$ pixels) on CRVD dataset \cite{RViDeNet} (denoted as ``$\pm16$ CRVD''). The misalignments between the reference frame and alternate frames are uniformly sampled in $[2,16]$ pixels.
  We combine Deformable Convolution alignment (DCN) in \cite{BDNet, RViDeNet} with our denoising network (denoted as ``Ours w/ DCN'').
  It is shown in Table~\ref{ab_align}, deformable convolution alignment only leads to marginal improvements over our method.
  We also evaluate the running times of the alignment modules on mobile processors. With the help of ARM Neon Intrinsic \cite{Arm}, our alignment is much faster than the DCN alignment.

\vspace{4pt}
\noindent \textbf{Multi-frame denoising network.} 
We first evaluate the effectiveness of sequential denoising on CRVD dataset (burst number $N=5$). When we remove the sequential denoising, the single denoising network would take all frames as inputs simultaneously (denoted as ``Ours-SQ1''). It is shown in Table~\ref{ab_seq_denoising} that the performance suffers from a drop of 0.33 dB PSNR when removing sequential denoising. When we use a network for spatial denoising on reference frame and another network for temporal denoising of the 4 alternate frames (denoted as ``Ours-SQ2''), the performance also drops by 0.24 dB PSNR. 
Furthermore, we test using a network for spatial denoising and two networks for sequential denoising and each network would handle temporal information of two neighboring alternate frames at a time (denoted as ``Ours-SQ3''). This design also causes a drop of $\sim$0.09 dB PSNR. In the above variants, we adjust the model size by changing the channel numbers so that different setups have similar FLOPs for fair comparison.

To demonstrate the advantages of the proposed multi-frequency aggregation,  we remove the multi-frequency aggregation and directly use the output $o_0$ as the final result (denoted as ``Ours w/o aggregation''). The network would be a little similar to SGN \cite{SGN_2019_ICCV}. It is shown in Table~\ref{ab_denoising_network} that removing multi-frequency aggregration causes a drop of about 0.26dB PSNR. This result demonstrates the advantages of the multi-frequency denoising. 
Then we test replacing the proposed multi-frequency aggragation with pointwise convolution (denoted as ``Ours w/ pointwise conv '') and attention-based fusion \cite{disneyasy} (denoted as ``Ours w/ attention''). When we adopt pointwise convolution for multi-frequency aggregation, it aggregates outputs of the three scales after $o_2, o_1$ are upsampled to the same size as $o_0$. It suffers a significant degradation of 0.72dB PSNR. 
As for attention-based fusion \cite{disneyasy},  we utilize an 8-layer convolutional network followed by a sigmoid layer to predict per-pixel aggregation weights between two neighboring frequencies.
It increases computational cost but still shows degraded performance of 0.15dB PSNR.

\section{Conclusions}
In this work, we proposed an efficient video denoising method via the improvements of three stages of the denoising framework noise prior integration, multi-frame alignment and multi-frame fusion.

Transforming raw images into a variance stabilization space can significantly reduce the model complexity without impacting its performance. From the perpective of on-chip running and efficiency, we combine classical image alignment and learning-based denoising to achieve comparable denoising performance with faster speed. As for multi-frame denoising, we introduce sequential denoising strategy and multi-frequency denoising to achieve efficient multiple-frame denoising. 

We have deployed these three stages on a commercial SoC. It demonstrates that our method can be employed for burst image denoising on mobile devices.

\begin{appendices}
\section{Noise modeling of CMOS Signals} \label{secA0}

We provide the detailed noise modeling of CMOS signals to obtain the relation between sensor gain and $\sigma_r, \sigma_s$. 
We define the observed intensity as $x$ and underlying true intensity as $x^{*}$. 
Following \cite{PMRID}, the raw signal is modeled as 
\begin{equation}
    \begin{aligned}
        x &\sim q_e \alpha \mathcal{P}\left(\frac{x^{*}}{q_e \alpha}\right) + \mathcal{N}(0, \alpha^2 \sigma_0^2 + \sigma_{adc}^2),
    \end{aligned}
\end{equation}
where $q_e$ is quantum efficiency factor, $\alpha$ is the sensor gain, $\sigma_0$ is the variance of read noise caused by sensor readout effects and $\sigma_{adc}$ is the variance of amplifier noise. 
Then we have:
\begin{equation}
    \begin{aligned}
        \sigma_s &= q_e a \\
        \sigma_r^2 &= \alpha^2\sigma_{0}^2 + \sigma_{adc}^2. 
    \end{aligned}
\end{equation}
For one fixed senor, $q_e$, $\sigma_{0}$, $\sigma_{adc}$ is unchanged. Then sensor gain $\alpha$ is the only factor to affect $\sigma_s, \sigma_r$.

\begin{figure}[!t]
\centering
\includegraphics[width=1.0\linewidth]{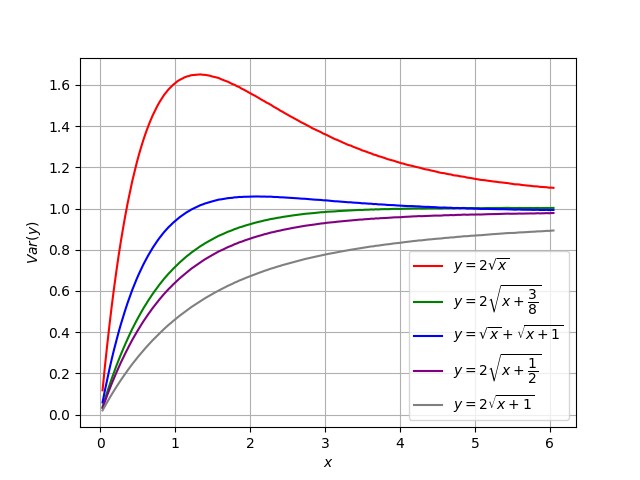}
\caption{Visualization of different variance stabilization transformations. $x$ is the mean of the signal in Poisson distribution. $Var(y)$ is the variance of the transformed signal by different transform function.
}
\label{fig:vis_transform}
\end{figure}

\section{Generalized verison of Freeman-Tukey Transformation}\label{secA1}
For For variable $x$ in Poisson distribution of the mean value $x^{*}$, the general form of variance stabilization transformation in root-type is 
\begin{equation}
    y = 2 \sqrt{x + c}. 
\end{equation}
The core problem of variance stabilization is to stabilize Poisson distribution to have unit variance. But no exact stabilization is possible \cite{transformation_analysis}. In practice, approximate transformations are generally used.
The mainstreaming transformations include $2\sqrt{x}$, $2\sqrt{x+1}$, $2\sqrt{x+\frac{1}{2}}$ \cite{1936Bart}, $2\sqrt{x+\frac{3}{8}}$ \cite{Anscombe} and $\sqrt{x}+\sqrt{x+1}$ \cite{tukey-freeman-transformation}. $\sqrt{x} + \sqrt{x+1}$ can be taken as the linear combination of two general forms with $c=0$ and $c=1$.
We visualize the variance of transformed y in Figure~\ref{fig:vis_transform}.
When the value $x$ is enough large, the variance of $2\sqrt{x+\frac{1}{2}}$ \cite{1936Bart}, $2\sqrt{x+\frac{3}{8}}$ \cite{Anscombe} and $\sqrt{x}+\sqrt{x+1}$ \cite{tukey-freeman-transformation} approach the unity. However, $\sqrt{x}+\sqrt{x+1}$ \cite{tukey-freeman-transformation} shows better approximation than other transformations when the mean value $x^{*}$ is close to zero. 
The SNR (signal-to-noise ratio) in dark areas is usually lower than that of other areas. Therefore, we seek the generalized version of Freeman-Tukey Transformation \cite{tukey-freeman-transformation} to handle Poisson-Gaussian distribution for raw denoising.

Firstly, we start from the transform of Poisson distribution. 
We define variable $x$ to be a Poisson variable of mean $m$. Its variance is $\text{Var}(x) = m$. We define $y$ to be the transformed x. Then we have $\text{Var}(y) \approx (\frac{dy}{dx})^2\text{Var}(x)$ based on \cite{delta_methods} and \cite{use_of_the_transformation}. The core problem of variance stabilization is stabilize Poisson distribution into unity variance. Hence we let $\text{Var}(y) = 1$ and obtain:
\begin{equation}
    \frac{dy}{dx} = \sqrt{\frac{\text{Var}(y)}{\text{Var}(x)}} = \frac{1}{\sqrt{m}}.
    \label{eq:at1}
\end{equation} 
For the general transform $y = 2\sqrt{x+c}$, we have 
\begin{equation}
    \frac{dy}{dx} = \frac{1}{\sqrt{x + c}}.
    \label{eq:at2}
\end{equation}
From Eqs. \eqref{eq:at1} and \eqref{eq:at2}, we obtain the approximation:
\begin{equation}
    m = x + c.
    \label{eq:pure_poisson}
\end{equation}

Secondly, we consider the transform of Poisson-Gaussian distribution. Similar to Eq. \eqref{eq:acute_x}, we define variable z as $z = x + \gamma$, where x is a Poisson variable of mean $m$ and $\gamma$ is a Gaussian variable of mean $g$ and standard deviation $\sigma$. The variance of transformed z is given by $\text{Var}(y) \approx (\frac{dy}{dx})^2\text{Var}(z)$ based on \cite{delta_methods} and \cite{use_of_the_transformation}.
 Similarly, we let $\text{Var}(y) = 1$ and obtain: 
\begin{equation}
    \frac{dy}{dz} = \sqrt{\frac{\text{Var}(y)}{\text{Var}(z)}} = \frac{1}{\sqrt{m + \sigma^2}}.
\end{equation}
We take the first-order approximation in \cite{GAT} to approximate the Gaussian distribution $\gamma \approx g$. From Eq. \eqref{eq:pure_poisson}, we have $m = z + c - g$. Thus we have: 
\begin{equation}
    \frac{dy}{dx} = \frac{1}{\sqrt{z + c + \sigma^2 - g}}.
    \label{eq:ftt_1}
\end{equation}
By integral of Eq. \eqref{eq:ftt_1}, we have the transformation y(z) for Poisson-Gaussian distribution:
\begin{equation}
    y(x) = 2\sqrt{z + c + \sigma^2 - g}.
    \label{eq:pg_vst}
\end{equation}

Finally, we move to the generalized version of Freeman-Tukey Transformation \cite{tukey-freeman-transformation}: $y = \sqrt{x} + \sqrt{x+1}$.  
From the Eq. \eqref{eq:pg_vst}, we generalize $2\sqrt{x}$ and $2\sqrt{x+1}$ respectively. 
By using linear combination of two generalized transformations
($c=0$ and $c=1$), we obtain the generalized version of Freeman-Tukey Transformation:
\begin{equation}
    y(x) = \sqrt{x + 1 + \sigma^2 - g} + \sqrt{x + \sigma^2 - g }.
\end{equation}

\begin{table}[!t]
\centering
\footnotesize
\setlength\tabcolsep{3pt}
\begin{tabular}{ccccc}
\hline
Algorithm & Variance Stabilization & Training loss  & Inverse & PSNR \\ \hline
GAT-1 & GAT (Anscombe)  & after inverse & unbiased & 44.60 \\
GAT-2 & GAT (Anscombe) & after inverse & algebraic & 44.60 \\
GAT-3 & GAT (Anscombe) & before inverse & unbiased & 44.50  \\
GAT-4 & GAT (Anscombe) & before inverse & algebraic& 44.63 \\ \hline
Ours-1 & Freeman-Tukey & after inverse & algebraic & 44.68 \\
Ours & Freeman-Tukey & before inverse & algebraic & 44.70\\ \hline

\end{tabular}
\caption{Ablation study of different inverse and different loss functions on CRVD dataset (burst number $N=5$).}
\label{ab_dinverse}
\end{table}

\section{Algebraic Inverse of transform}
It is known that algebraic inverse is usually avoided due to bias in previous methods \cite{GAT}.
However the bias is already handled when we calculate the loss in the space of variance stabilization. 
Moreover, algebraic inverse can be used for both Anscombe transformation \cite{Anscombe,GAT} and Freeman-Tukey transformation \cite{tukey-freeman-transformation} in our framework. 

Let $x$ and $x^{*}$ denote noisy signal and clean signal, respectively. The transform (Anscombe transform or Freeman-Tukey transform) is denoted as $f$ and the algebraic transform is denoted as $f^{-1}$. The bias is produced by the nonlinearity of the transformation $f$. We calculate the loss in the variance stabilization space. The denoising network would learn the mapping from $f(x)$ to $f(x^{*})$ directly. Therefore, the bias is already handled when the denoising output approximates $f(x^{*})$.

We further conduct experiments on CRVD dataset (burst number $N=5$) to compare algebraic inverse and exact unbiased inverse under different training settings. The results are shown in Table~\ref{ab_dinverse}.
We first training with Generalization Anscombe transformation (GAT) \cite{GAT} and calculate the loss function before the inverse. Then we test the model with algebraic inverse (denoted as ``GAT-4'') and exact unbiased inverse (denoted as ``GAT-3''). It is shown that algebraic inverse outperforms the exact unbiased inverse \cite{optimal_invserse} by 0.13 dB PSNR, which demonstrates that the bias is handled in calculating loss before inverse.
Then we train with GAT with algebraic inverse (denoted as ``GAT-2'') and optimal inverse (denoted as ``GAT-1'') and calculate the loss function after the inverse.
In Table~\ref{ab_dinverse}, it can be observed that both two inverses show the same performance (44.60 dB PSNR) but are 0.03 dB PSNR lower than calculating the loss before inverse. It might be because the bias produced in the space of variance stabilization becomes more complicated after the non-linear inverse transformation. Handling the bias before inverse is more direct. The same phenomenon can also observed in the Freeman-Tukey transformation (``Ours-1'' VS ``Ours'').

\begin{table}[!t]
\centering
\setlength\tabcolsep{3pt}
\begin{tabular}{cccc}
\hline
Input order  & Reverse & Shuffle & Keep \\ \hline
PSNR (dB) & 44.63 & 44.67 & 44.70 \\ \hline
\end{tabular}
\caption{Ablation study of different input orders of alternate frames on CRVD dataset (burst number $N=5$).}
\label{ab_input_order}
\end{table}

\begin{table}[!t]
\centering
\setlength\tabcolsep{3pt}
\begin{tabular}{cccc}
\hline
Network weights & Sharing & Specializing \\ \hline
PSNR (dB) & 44.44 & 44.70 \\ \hline
\end{tabular}
\caption{Ablation study of using specialized or shared-weights networks on CRVD dataset (burst number $N=5$).}
\label{ab_specializing}
\end{table}

\begin{table}[!t]
\centering
\setlength\tabcolsep{3pt}
\begin{tabular}{cccc}
\hline
\# Scales  & $s=2$ & $s=3$ & $s=4$ \\ \hline
PSNR (dB) & 44.49 & 44.70 & 44.71 \\ 
Params. (M) & 1.06 & 1.57 & 2.10 \\ \hline
\end{tabular}
\caption{Ablation study of using different numbers of frequencies in the denoising network on CRVD dataset (burst number $N=5$).}
\label{ab_low_freq}
\end{table}

\section{More Ablation of Denoising Network}
\textbf{Input order of alternate frames} We conduct experiments on CRVD dataset \cite{RViDeNet} (burst number $N=5$) to compare three input orders: a) preserving the temporal order of an input burst (denoted as ``Keep''), b) shuffling the burst order randomly (denoted as ``Shuffle''), and c) reversing the burst order (denoted as ``Reverse''). In training and testing, 4 alternate frames are re-arranged following the same ordering strategies.
It is shown in Table~\ref{ab_input_order} that training with preserving the temporal order achieve the best performance of 44.70 dB PSNR, which slightly outperforms random shuffling by 0.03 dB PSNR.
Furthermore, reversing the temporal order achieve the worst performance of 44.63 dB PSNR, which suffers a drop of 0.07dB PSNR. It can be observed that preserving the temporal order is helpful in sequential denoising.

\noindent\textbf{Specializing the network weights} 
In our denoising network $S$, we have a series of sub-networks for sequential denoising. 
For burst denoising on CRVD dataset \cite{RViDeNet} (burst number $N=5$), $S_0$ is for spatially denoising of reference frame and $S_1,S_2,S_3,S_4$ are for sequential denoising of the 4 alternate frames. 
We conduct experiments on CRVD dataset (burst number $N=5$) to compare $S_i$ with different weights (denoted as ``specializing'') and $S_i$ with shared weights (denoted as ``sharing''). It shown in Table~\ref{ab_specializing} that using shared weights of $S_1,S_2,S_3,S_4$ just achieves 44.44 dB PSNR, which has a drop of 0.26 dB PSNR compared with specializing each $S_i$ (44.70 dB PSNR).

\noindent\textbf{Different Scales in denoising backbone} We conduct experiments on CRVD dataset (burst number $N=5$) to explore using different scales (frequencies). We define the number of scales (frequencies) as $s$. When $s=4$, we use four frequencies ($m_0,m_1,m_2,m_3$) to achieve multi-frequency denoising. 
It can be observed in Table~\ref{ab_low_freq} that using two frequencies achieves 44.49 dB PSNR, which is a drop of 0.21dB compared with using three frequencies (44.70 dB PSNR). 
But when we use four scales (frequencies), the denoising performance is 44.71 dB PSNR and just outperform using three frequencies by only 0.01 dB PSNR but its model size increases from 1.57M to 2.10M.

\end{appendices}

\bibliographystyle{plain}
\bibliography{egbib}

\end{document}